# A Stochastic Model of Active Cyber Defense Dynamics

Shouhuai Xu[†], Wenlian Lu[‡⋆], and Hualun Li[⋆]

[†] Department of Computer Science, University of Texas at San Antonio
San Antonio, Texas 78249, USA

[‡] Department of Computer Science, University of Warwick
Coventry CV4 7AL, UK

[⋆] School of Mathematical Sciences, Fudan University
Shanghai, P. R. China, 200433

Email: `shxu@cs.utsa.edu, wenlian@fudan.edu.cn, 09210180042@fudan.edu.cn`

August 5, 2013

**Abstract**

The concept of active cyber defense has been proposed for years. However, there are no mathematical models for characterizing the effectiveness of active cyber defense. In this paper, we fill the void by proposing a novel Markov process model that is native to the interaction between cyber attack and active cyber defense. Unfortunately, the native Markov process model cannot be tackled by the techniques we are aware of. We therefore simplify, via mean-field approximation, the Markov process model as a Dynamic System model that is amenable to analysis. This allows us to derive a set of valuable analytical results that characterize the effectiveness of four types of active cyber defense dynamics. Simulations show that the analytical results are inherent to the native Markov process model, and therefore justify the validity of the Dynamic System model. We also discuss the side-effect of the mean-field approximation and its implications.

## 1 Introduction

The concept of *active cyber defense* (e.g., using the so-called "white" or "good" worms to identify and fight/kill the malicious ones) has been proposed for years. However, the exploration has primarily focused on legal and policy issues [1, 2, 3, 4, 5, 6, 7, 8]. On the other hand, active cyber defense has already happened in some sense (e.g., the `Welchia` worm attempted to "kick out" the Blaster worm from the infected computers [4, 9]), and full-fledged active cyber defense is seemingly inevitable in the near future [6, 10, 11]. It is therefore more imperative than ever to systematically characterize the effectiveness of active cyber defense. In this paper, we initiate the theoretical study on this perspective of cyber security, with emphasis on addressing the following basic question: *How effective is active cyber defense?* Such characterization studies not only will deepen our understanding of active cyber defense, but also will help real-life decision-making (e.g., when to launch active cyber defense?) and even policy-making (e.g., whether or not to launch active cyber defense?).

### 1.1 Our Contributions

We formulate, to the best of our knowledge, the first mathematical model for characterizing the effectiveness of active cyber defense. The interaction between cyber attack and active cyber defense can be naturally modeled as a Markov process (Section 2). Unfortunately, we do not know how to tackle the native Markov process analytically because all the techniques we are aware of do not appear to be applicable (see Section 1.2 for discussions). We therefore simplify, via the mean-field approximation, the native Markov process model as a Dynamic System model that is amenable to analysis. In the Dynamic System model, we obtain a set of analytical results (Sections 4-6). We then use simulations to validate the accuracy of the Dynamic System model (Section 8). Simulations show that the analytical results derived from the Dynamic System model are inherent to the native Markov process model, and that the accuracy of the Dynamic System model, in terms of *dynamics accuracy* and *threshold accuracy* (which will be specified in Section 8), increases with the average node degree. Moreover, the analytical results lead to various insights, with some highlighted (informally) as follows:

- If neither the defender nor the attacker is superior to, or more advanced than, its opponent in terms of cyber combat power (Types I-II dynamics with a certain threshold), the effectiveness of active cyber defense will depend on (in some quantitative fashion we derive): (i) the attack-defense network structure, (ii) the initial security state of the attack-defense network, (iii) the attacker's and defender's combat-power, and (iv) the attacker/defender strategy. We also characterize the benefit to *strategic* attacker/defender that initially "occupies" the large-degree nodes. Specifically, we show: (i) when the attack-defense network structures are Erdös-Rényi (ER) random graphs, a strategic defender/attacker does not gain significant benefit; (ii) when the attack-defense network structures are power-law graphs, a strategic defender/attacker gains significant benefit.[1] Moreover, we obtain the following quantitative result: The benefit to strategic defender is maximized for the sub-class of power-law graphs with exponent $\gamma = 2$. These are described in Sections 4-5.

- If the defender is superior to (or more advanced than) the attacker in terms of cyber combat power (Type III dynamics), the defender can always use active cyber defense to *automatically* "clean up" (i.e., cure) the entire network, regardless of the attack-defense network structure and no matter whether the attacker is strategic or not. This suggests that cyber superiority could serve as an effective deterrence, and can be seen as a consequence due to the lack of a certain threshold in the combat power function. The explorations of Type III dynamics and its dual (i.e., Type IV dynamics) are described in Section 6.

- As discussed in Section 7, active cyber defense can eliminate the asymmetry that is an inherent weakness of reactive cyber defense, where the defender runs "anti-virus software"-like tools on each computer to detect and cure infections (which are caused by that the attacks/malwares penetrated the perimeter defense such as Firewalls). The cause of the asymmetry is that when the defense is reactive, the attack effect is automatically amplified by the network (a kind of "network effect").

We stress that the focus of the present paper is to characterize how effective active cyber defense is. This means that we should not make any significant restrictions on the parameter regimes and network structures. One important research problem, which is orthogonal to our focus and is not addressed in the present paper, is how to extract the model parameters and the attack-defense network structure for a given cyber system. In principle, the model parameters can be obtained by analyzing the strength and weakness of the attack and defense tools ("what if" analysis can be used in the absence of sufficient data), and/or by conducting experiments (in lieu of physical experiments) to observe the outcome of experimental cyber combats. The network structure can be derived from the cyber system configurations and security policies, which may restrict which computers can directly communicate with which other computers. The characterization results presented in this paper accommodate a large class of parameter and structure scenarios.

---

[1] These results are reminiscent of, and in parallel to, the *connectivity-based* robustness characterizations of ER and power-law graphs [12], which is however a different perspective from ours because the attacker in our model aims to compromise as many nodes as possible but does *not* delete any (of the compromised) nodes.

## 1.2 Related Work

We classify the related prior work based on two perspectives: one is centered on the problem that is under investigation, and the other is centered on the technique that is exploited to tackle the problem under investigation.

From the perspective of the problem under investigation, we note that all existing studies in both the mathematics literature and the physics literature are geared toward, in the terms of the present paper, characterizing the outcome of *reactive* defense under various parameter conditions (see for example [13, 14, 15, 16, 17, 18, 19, 20, 21, 22, 23] and the references there in). These studies substantially generalize the pioneering work of Kephart and White [24, 25], which was based on *homogeneous* epidemic models in biological systems [26, 27, 28]. For example, even for the very recent work [21], which studies the attack-defense dynamics between one defender and multiple attackers that fight against each other as well, the defense is still *reactive*. In contrast, the present paper introduces a new research problem, namely characterizing the outcome of *active* defense under various model parameter conditions (include the graph/network structure). To the best of our knowledge, we are the first to study the active cyber defense problem mathematically, despite that the technical practice of active cyber defense has been discussed for years [1, 2, 3, 4, 5, 6, 7, 8]. This is so even though our active cyber defense model is reminiscent of the *voter model* (see, for example, [29, 30, 31, 32, 33]), where each node can adopt the state of one of its random neighbors at each time step. However, the voter model corresponds to the special case of our active cyber defense model with *linear* combat-power functions (the concept of combat-power functions will be introduced later). In contrast, we study general *nonlinear* combat-power functions, which explain why the techniques for analyzing the voter model cannot tackle our active cyber defense model (see Section 2.2 for further discussions). Finally, it is worth mentioning that active cyber defense is different from automatic patching [34] because the attacker may have already compromised many computers, and that our active cyber defense model is different from the Moran process [35, 36], which considers the mutation dynamics of homogeneous nodes.

From the perspective of the techniques that are exploited to tackle the epidemic problem with network structures [13, 14, 15, 16, 17, 18, 19, 20, 21, 22, 23], there are mainly two approaches. The first approach is to use the mean-field approximation (e.g., [37]). Our Dynamic System model is also based on mean-field approximation of a native stochastic process model. Mean-field approximation is a plausible first step in studying problems such as the stochastic active cyber defense process we introduce in the present paper. Nevertheless, we empirically characterize the accuracy of the mean-field approximations.

The second approach is to directly tackle the native processes that take place on network structures. This approach is more rigorous than the mean-field approximation approach, but is often pursued after establishing some understandings based on the mean-field approach. This approach is valuable not only because it can derive rigorous results, but also because it can (in)validate some results obtained via the mean-field models. For example, Ball et al. [19, 20] study the threshold behavior and the final outcome of the SIR (susceptible-infectious-removed) epidemic process on random networks with clusters (communities). They consider the SIR epidemic process in two steps: the SIR epidemic spreading within the clusters (local spreading) and then the SIR epidemic spreading cross the clusters. For their studies, it is reasonable to use the Branching process approximation because they only need to consider the case of small initial infections (i.e., early stage of epidemic spreading) and because the notion of *offspring generation* is well-defined in SIR models. They derive a rigorous central limit theorem under certain conditions. In another line of investigations, Berger et al. [15] investigate the SIS (susceptible-infectious-susceptible) Contact process [38] on random graphs that are generated via preferential attachment [39]. Their rigorous study confirms the threshold result of Pastor-Satorras and Vespignani [37] obtained via the mean-field approximation, namely that the epidemic threshold of scale-free networks is 0. Chatterjee and Durrett [18] further study both SIR and SIS models on random graphs with the power-law degree distributions. Improving upon some results in [18], Mountford et al. [22] show that the epidemic extinction time for the Contact process on power-law random graphs grows exponentially in the number of nodes, and Mountford et al. [23] obtain bounds for the density of infected nodes.

The rest of the paper is organized as follows. In Section 2, we present the native Markov process model and then show how to simplify it as a Dynamic System model that is amenable to analysis. In Section 3, we briefly review some background knowledge. In Sections 4-6, we characterize four types of active cyber defense dynamics. In Section 7, we explain why active cyber defense can eliminate an inherent weakness of reactive cyber defense. In Section 8, we use simulations to show that the analytical results derived from the Dynamic System model are inherent to the native Markov process model. In Section 9, we conclude the paper with future research directions. Lengthy proofs are deferred to the Appendix.

## 2 Active Cyber Defense Model

A cyber system consists of networked computers/nodes of finite populations. A computer has two states: compromised or secure (i.e., vulnerable but not compromised). We may say that a compromised computer is "occupied" by the adversary/attacker, and a secure computer is "occupied" by the defender. The adversary can compromise a computer by exploiting its (e.g., zero-day or unpatched) vulnerabilities. Attacks are malware-like, meaning that the compromised computers can attack the vulnerable computers in an epidemic-spreading fashion. With active cyber defense, the defender can use "good worm"-like mechanisms to spread in networks (as the malicious worms do) to identify and "clean up" the compromised computers.

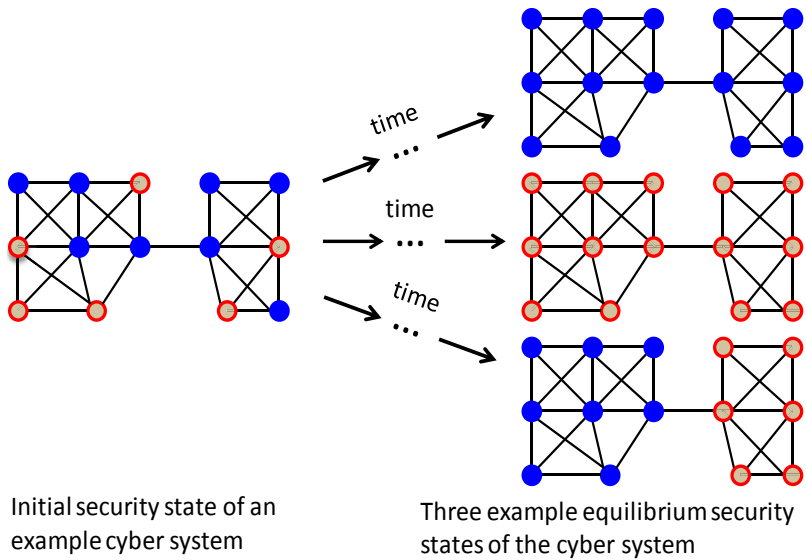

Initial security state of an example cyber system

Three example equilibrium security states of the cyber system

Figure 1: Illustration of cyber security state evolution under active cyber defense, where the same initial state may evolve, under different conditions, toward one of the three example equilibrium states — all nodes are secure (blue dots); all nodes are compromised (red dots); some nodes are secure. The core research issue is to characterize how the initial state, network topology, parameters and attacker/defender strategies would govern the evolution.

The interaction between cyber attack and active cyber defense formulates an attack-defense interaction structure, a graph topology that represents how the compromised nodes attack the secure nodes and how the secure nodes use active cyber defense to clean up the compromised nodes. We say a defender (attacker) is *strategic* if it initially occupies the large-degree nodes in the graph with higher probabilities. The attack-defense interaction leads to the evolution of cyber security state of the entire cyber system. We illustrate the state evolution in Figure 1, where a blue dot means "secure" and a red dot means "compromised." As shown in Figure 1, the state evolution can exhibit rich phenomena (e.g., the existence of multiple kinds of equilibria). At a high level, the research objective is to characterize how the evolution is governed by the initial state, graph topology, parameters and attacker/defender strategies. The characterization will allow us to answer some basic questions such as: under what conditions the cyber security state evolves toward the all-blue equilibrium?

## 2.1 The Native Markov Process Model

Formally, cyber attack-defense takes place over a finite network/graph structure $G = (V, E)$, where $V = \{1, 2, \ldots, n\}$ is the set of nodes/computers and $E$ is the set of edges/arcs with $(u, u) \notin E$ (i.e., there are no self-loops in the setting of the problem). At any point in time, a node $v \in V$ is in one of two states: blue, meaning that it is *secure* (i.e., vulnerable but not compromised by the attacker); red, meaning that it is *compromised* by the attacker. Node $v$'s state changes because of some $u$, where $(u, v) \in E$. Note that $(u, u) \notin E$ because a secure node will not clean up itself and a compromised node will not attack itself. Since our study applies to both undirected and directed graphs, we focus on undirected graphs while mentioning the difference when the need arises. We do not make any significant restrictions on $G$ because in real-life $G$ can have any topology. This has become a standard practice in characterization studies of cyber security (see, for example, [13, 14, 16, 17, 21]).

The state of node $v \in V$ at time $t$ is a random variable $\xi_v(t) \in \{0, 1\}$:

$$\xi_v(t) = \begin{cases} 1 & v \in V \text{ is blue at time } t \\ 0 & v \in V \text{ is red at time } t. \end{cases}$$

Correspondingly, we define

$$B_v(t) = \mathsf{P}(\xi_v(t) = 1) \text{ and } R_v(t) = \mathsf{P}(\xi_v(t) = 0).$$

Denote by $\tilde{\theta}_{v,BR}(t)$ the rate at which $v$'s state changes from blue to red at time $t$, which is a random variable because it depends on the states of $v$'s neighbors. Similarly, denote by $\tilde{\theta}_{v,RB}(t)$ the random rate at which $v$'s state changes from red to blue at time $t$. The state evolution of $v \in V$ is naturally described as a Markov process (dubbed "Markov process model" or "Markov model" for reference purpose) with the following transition probabilities:

$$\mathsf{P}(\xi_v(t + \Delta t) = 1 | \xi_v(t)) = \begin{cases} \Delta t \cdot \tilde{\theta}_{v,RB}(t) + o(\Delta t) & \xi_v(t) = 0 \\ 1 - \Delta t \cdot \tilde{\theta}_{v,BR}(t) + o(\Delta t) & \xi_v(t) = 1 \end{cases} \quad (1)$$

and

$$\mathsf{P}(\xi_v(t + \Delta t) = 0 | \xi_v(t)) = \begin{cases} \Delta t \cdot \tilde{\theta}_{v,BR}(t) + o(\Delta t) & \xi_v(t) = 1 \\ 1 - \Delta t \cdot \tilde{\theta}_{v,RB}(t) + o(\Delta t) & \xi_v(t) = 0 \end{cases} \quad (2)$$

as $\Delta t \to 0$. Denote by $N_v = \{u \in V : (u, v) \in E\}$ the set of neighbors of node $v \in V$. Since the random rates $\tilde{\theta}_{v,RB}(t)$ and $\tilde{\theta}_{v,BR}(t)$ are naturally determined by the random states of node $v$'s neighbors, we use deterministic but possibly nonlinear functions $f_{RB}(\cdot) : \mathbb{R} \to [0, 1]$ and $f_{BR}(\cdot) : \mathbb{R} \to [0, 1]$ to define respectively the random rates $\tilde{\theta}_{v,RB}(t)$ and $\tilde{\theta}_{v,BR}(t)$, as follows:

$$\tilde{\theta}_{v,RB}(t) = f_{RB}\left(\frac{1}{\deg(v)} \sum_{u \in N_v} \xi_u(t)\right) \text{ and } \tilde{\theta}_{v,BR}(t) = f_{BR}\left(\frac{1}{\deg(v)} \sum_{u \in N_v} (1 - \xi_u(t))\right).$$

We call $f_{RB}(\cdot)$ and $f_{BR}(\cdot)$ the *combat-power* functions because they abstract the attacker's and defender's combat capabilities.

At this point, we do not know how to tackle the above native Markov process model. One may note that the above combat-power functions are reminiscent of the so-called Voter model [29], where a node changes its opinion (or state) to the opinion of one random neighbor according to a *fixed-rate* Poisson process. This allows the model to be transformed into a *dual* process that works backward in time and becomes a random walk [29], which makes it tractable. In contrast, in our model a node changes its state according to a rate that is not fixed but instead *nonlinearly* dependent up on the states of its neighbors. The nonlinearity prevents us from transforming our native Markov process model into a Random Walk model, meaning that the technique used in [29] cannot solve the problem we encounter. This nonlinearity-caused difficulty suggests us to simplify/approximate the native Markov process model as a tractable Dynamic System model.

## 2.2 Simplifying the Markov Process Model as Dynamic System Model

Now we show how to simplify the native Markov process model into a tractable Dynamic System model via the mean-field approximation. From Eq. (1), we have, for $v \in V$,

$$B_v(t + \Delta t) = \Delta t \cdot \tilde{\theta}_{v,RB}(t) \cdot R_v(t) + (1 - \Delta t \cdot \tilde{\theta}_{v,BR}(t))B_v(t) + o(\Delta t),$$

which can be rewritten as:

$$\frac{B_v(t + \Delta t) - B_v(t)}{\Delta t} = \tilde{\theta}_{v,RB}(t) \cdot R_v(t) - \tilde{\theta}_{v,BR}(t) \cdot B_v(t) + o(\Delta t).$$

Similarly, from Eq. (2) we can derive for all $v \in V$:

$$\frac{R_v(t + \Delta t) - R_v(t)}{\Delta t} = \tilde{\theta}_{v,BR}(t) \cdot B_v(t) - \tilde{\theta}_{v,RB}(t) \cdot R_v(t) + o(\Delta t).$$

By letting $\Delta t \to 0$, we have for all $v \in V$:

$$\begin{cases} \frac{d}{dt}B_v(t) = \tilde{\theta}_{v,RB}(t) \cdot R_v(t) - \tilde{\theta}_{v,BR}(t) \cdot B_v(t) \\ \frac{d}{dt}R_v(t) = \tilde{\theta}_{v,BR}(t) \cdot B_v(t) - \tilde{\theta}_{v,RB}(t) \cdot R_v(t). \end{cases} \quad (3)$$

Note that

$$\mathsf{E}\left(\tilde{\theta}_{v,RB}(t)\right) = \mathsf{E}\left(f_{RB}\left(\frac{1}{\deg(v)} \sum_{u \in N_v} \xi_u(t)\right)\right).$$

By the idea of mean-field approximation, we can move the expectation inside the combat-power function, and replace the mean of random rate $\tilde{\theta}_{v,RB}(t)$, denoted by $\theta_{v,RB}(t)$, with the following term:

$$f_{RB}\left(\frac{1}{\deg(v)} \sum_{u \in N_v} \mathsf{E}\left[\xi_u(t)\right]\right) = f_{RB}\left(\frac{1}{\deg(v)} \sum_{u \in N_v} B_u(t)\right).$$

We can treat $\tilde{\theta}_{v,BR}(t)$ analogously. As a result, we obtain the mean state-transition probability $\theta_{v,RB}(t)$ and $\theta_{v,BR}(t)$ as:

$$\theta_{v,RB}(t) = f_{RB}\left(\frac{1}{\deg(v)} \sum_{u \in N_v} B_u(t)\right) \quad \text{and} \quad \theta_{v,BR}(t) = f_{BR}\left(\frac{1}{\deg(v)} \sum_{u \in N_v} R_u(t)\right).$$

Therefore, Eq. (3) becomes the following Dynamic System model for all $v \in V$:

$$\begin{cases} \frac{d}{dt} B_v(t) = \theta_{v,RB}(t) \cdot R_v(t) - \theta_{v,BR}(t) \cdot B_v(t) \\ \frac{d}{dt} R_v(t) = \theta_{v,BR}(t) \cdot B_v(t) - \theta_{v,RB}(t) \cdot R_v(t). \end{cases} \quad (4)$$

Note that the Dynamic System model for all $v \in V$ encodes the graph topology via parameters $\theta_{v,BR}(t)$ and $\theta_{v,RB}(t)$, which encode the information about node $v$'s neighborhood (including the states of node $v$'s neighbors). The corresponding *state-transition diagram* for any node $v \in V$ is depicted in Figure 2.

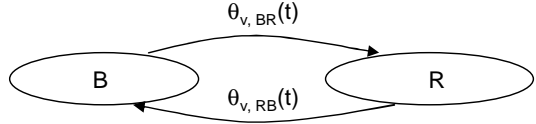

Figure 2: State-transition diagram of a single node $v \in V$ (B: blue; R: red)

## 2.3 Instantiating the Dynamic System Model via Specific Combat-Power Functions

Recall that combat-power function $f_{RB}(\cdot)$ abstracts the defender's power against the attacker. It should satisfy the following properties: (i) $f_{RB}(0) = 0$; (ii) $f_{RB}(1) = 1$; (iii) $f_{RB}(\cdot)$ increases monotonically. This is intuitive because the more blue nodes surrounding a red node, the greater the chance the red node will become blue (because of the active defense launched by the blue nodes). In this paper, we consider four types of $f_{RB}(\cdot)$ with examples depicted in Figure 3, where the first two types of $f_{RB}(\cdot)$ have an inherent threshold while the others don't.

- Type I: For a given threshold $\sigma \in (0,1)$, we define

$$\theta_{v,RB}(t) = f_{RB}\left(\frac{1}{\deg(v)} \sum_{u \in N_v} B_u(t)\right) = \begin{cases} 1 & \frac{1}{\deg(v)} \sum_{u \in N_v} B_u(t) > \sigma \\ 0 & \frac{1}{\deg(v)} \sum_{u \in N_v} B_u(t) < \sigma \\ 1/2 & \text{otherwise.} \end{cases} \quad (5)$$

Intuitively, the defender is more powerful than the attacker when $\sigma < 1/2$, less powerful than the attacker when $\sigma > 1/2$, and equally powerful as the attacker when $\sigma = 1/2$.

- Type II: For a given threshold $\tau \in (0,1)$, we define: $f_{RB}(x)$ is convex and $f_{RB}(x) < x$ for $x \in [0, \tau)$; $f_{RB}(x)$ is concave and $f_{RB}(x) > x$ for $x \in (\tau, 1]$; $f_{RB}(x) = x$ for $x = \tau$, $f_{RB}(0) = 0$, and $f_{RB}(1) = 1$. Moreover, $f_{RB}(\cdot)$ is increasing and continuous in intervals $[0, \tau)$ and $(\tau, 1]$. This type of functions is known as "sigmoid" functions. Intuitively, the defender is more powerful than the attacker when $\tau < 1/2$, less powerful than the attacker when $\tau > 1/2$, and equally powerful as the attacker when $\tau = 1/2$.

- Type III: $f_{RB}(\cdot)$ is concave, continuous and increasing in $[0,1]$, $f_{RB}(x) > x$ for $s \in (0,1)$ and $f_{RB}(0) = 0$, $f_{RB}(1) = 1$. Intuitively, the defender is more advanced than the attacker (i.e., the defender has cyber combat superiority).

- Type IV: $f_{RB}(\cdot)$ is convex, continuous and increasing in $[0,1]$, $f_{RB}(x) < x$ for $x \in (0,1)$ and $f_{RB}(0) = 0$, $f_{RB}(1) = 1$. Intuitively, the defender is less advanced than the attacker. Note that Type IV $f_{RB}(\cdot)$ is dual to Type III $f_{RB}(\cdot)$.

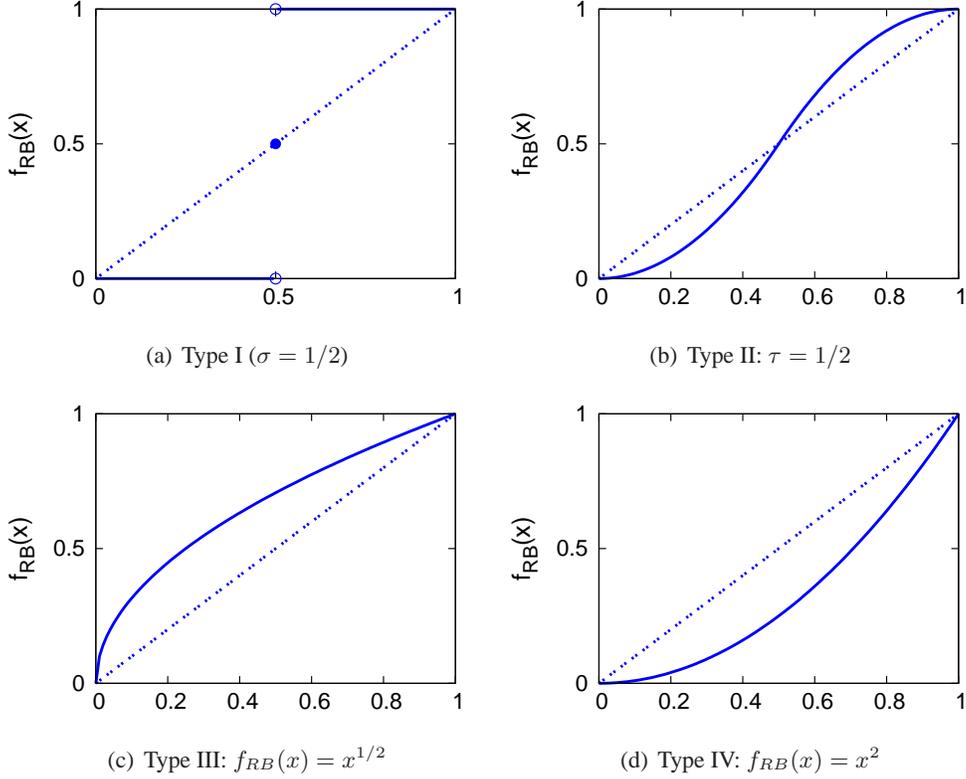

Figure 3: $f_{RB}(\cdot)$ examples (Type II: $f_{RB}(x) = 2x^2$ for $x \in [0, 0.5]$, $f_{RB}(x) = -2x^2 + 4x - 1$ for $x \in [0.5, 1]$)

Based on the above 4 types of combat-power functions, we focus on the 4 types of combat-function combinations that satisfy

$$\theta_{v,BR}(t) = 1 - \theta_{v,RB}(t). \tag{6}$$

By combining Eqs. (4) and (6), we obtain the following master equation for a *single* node $v \in V$:

$$\begin{aligned}\frac{d}{dt}B_v(t) &= \theta_{v,RB}(t)(1 - B_v(t)) - \theta_{v,BR}(t)B_v(t) \\ &= \theta_{v,RB}(t) - B_v(t).\end{aligned} \tag{7}$$

The research task is to characterize Types I-IV dynamics, namely the dynamics of master equation (7) with Types I-IV combat-power functions, respectively. For example, for Type I combat-power function, we have

$$\theta_{v,BR}(t) = f_{BR}\left(\frac{1}{\deg(v)}\sum_{u \in N_v}[1 - B_u(t)]\right) = \begin{cases} 1 & \frac{1}{\deg(v)}\sum_{u \in N_v}[1 - B_u(t)] > 1 - \sigma \\ 0 & \frac{1}{\deg(v)}\sum_{u \in N_v}[1 - B_u(t)] < 1 - \sigma \\ 1/2 & \text{otherwise.} \end{cases} \tag{8}$$

We want to characterize, among other things, the roles of the *thresholds* specified in Types I-II dynamics, and the consequences due to the lack of such thresholds in Types III-IV dynamics.

**Summary of notations**

Let $\mathbb{R}$ be the set of real numbers, $\mathsf{P}(\cdot)$, $\mathsf{E}(\cdot)$, $\mathsf{Var}(\cdot)$ be the probability, expectation, and variance functions, respectively. Other major notations are summarized in the following table.

| | |
|---|---|
| $G = (V, E)$ | graph/network that abstracts a cyber system from a cyber security perspective, where $|V| = n$ |
| $N_v$ | $N_v = \{u \in V : (u, v) \in E\}$ |
| $\deg(v)$ | $v$'s (in-)degree, $\deg(v) = |N_v|$ |
| $\gamma$ | power-law exponent, $\mathsf{P}(\deg(v) = k) \propto k^{-\gamma}$ |
| $\sigma, \tau$ | indicator of defender's relative combat-power in Types I and II dynamics, respectively |
| $\xi_v(t)$ | state of node $v$ at time $t$: blue (i.e., 1 or "secure") and red (i.e., 0 or "compromised") |
| $B_v(t)$ | probability $v \in V$ is blue at time $t$ |
| $R_v(t)$ | probability $v \in V$ is red at time $t$ |
| $\alpha$ | $\alpha = \frac{1}{n}\sum_{v \in V} B_v(0)$, namely the average fraction of blue nodes at time $t = 0$ |
| $S$ | random set of blue nodes at time $t = 0$ |
| $\theta_{v,BR}(t)$ | probability that node $v$'s state changes from blue to red at time $t$ |
| $\theta_{v,RB}(t)$ | probability that node $v$'s state changes from red to blue at time $t$ |

## 3 Preliminaries

**Arbitrary Networks**

By "arbitrary network" we mean a *given* network $G = (V, E)$ that may or may not have a special structure/topology of interest. Most analytical results in this paper are derived from dynamic systems that take place on arbitrary networks. In general, such results are often independent of the statistical properties of the networks (e.g., the degree distribution).

In order to show the existence of the third kind of equilibrium illustrated in Figure 1 (i.e., some nodes in blue color and the other nodes in red color), we also consider a given network that has a cluster (or community) structure. A network $G = (V, E)$ has a clustered structure of $V_1, V_2, \ldots, N_K$ if $\bigcup_i V_i = V$, $V_i \cap V_j = \emptyset$ for all $i \neq j$, and the nodes belonging to $V_i$ are better connected than the nodes crossing $V_i$ and $V_j$ for any $i \neq j$. More specifically, the special phenomenon is related to the minimum node expansion in cluster $V_k$ for $1 \leq k \leq K$, which is defined as

$$\beta_k = \inf_{v \in V_k} \frac{|N_v \cap V_k|}{\deg(v)}, \quad \text{where } N_v = \{u : (u, v) \in E\}. \tag{9}$$

**Generalized Random Graphs**

In order to characterize the benefit to the *strategic* defender who initially occupies the large-degree nodes with greater probabilities (a scenario that is often difficult to analyze), we propose to use the generalized random graph model [40]. This means that the result is applicable to a class of random networks (which however include the Erdös-Rényi (ER) random graphs and power-law random graphs [40]), rather than arbitrary networks; this slight restriction is compensated with some valuable analytical results. (Characterizing the benefit to strategic defender in *arbitrary* networks is left as an open problem.)

In the generalized random graph model, we are given an expected (in-)degree sequence $(d_1(n), \ldots, d_n(n))$ that defines a family of graphs. Let $d_{\min}(n) = \min\{d_j(n) : 1 \leq j \leq n\}$ and $d_{\max}(n) = \max\{d_j(n) : 1 \leq j \leq n\}$. A random graph instance $G(n) = (V(n), E(n))$ can be obtained by linking each pair of nodes $(u, v)$ with probability

$$p_{vu}(n) = \frac{d_u(n)d_v(n)}{\sum_{k=1}^n d_k(n)} \tag{10}$$

independent of the others [40], where $0 \leq p_{vu}(n) \leq 1$ under the assumption $(d_{\max}(n))^2 \leq \sum_{k=1}^n d_k(n)$.

For simplifying the analysis, we allow self-links while noting that our result can be adapted to accommodate that there are no self-links. In order to attain deeper insights, we will consider two instantiations of the generalized random graph model, namely the classic Erdös-Rényi (ER) random graphs with $d_1(n) = \ldots = d_n(n)$ or edge probability $p = d_1(n)/n$, and the ubiquitous power-law random graphs with $\frac{\#\{v \in V : \deg(v) = k\}}{\#V} \propto k^{-\gamma}$ for some $\gamma > 0$. Note that $\gamma$ does not need to be greater than 1 because $d_{\max}(n)$ is finite.

Note that complete graphs are a special case of arbitrary networks and of generalized random graphs. Since the theorems we present below hold either for arbitrary networks or for generalized random graphs, they automatically apply to complete graphs.

## 4 Characterizing Type I Active Cyber Defense Dynamics

In this section, we first characterize Type I active cyber defense dynamics with *non-strategic* defender in arbitrary networks, where the initial occupation probability $B_v(0)$ is identical to all nodes. We then investigate the more difficult case of *strategic* defender with degree-dependent $B_v(0) \propto \deg(v)$ in the generalized random graph model, where the defender initially occupies the large-degree nodes with higher probabilities (i.e., the large-degree nodes are appropriately better protected).

### 4.1 Characterizing Type I Dynamics with Non-Strategic Defender

Type I dynamics with non-strategic defender is characterized through Theorems 1-3. The characterizations include the conditions under which the defender can or cannot use active cyber defense to automatically clean up the entire network, and a method for deciding whether an equilibrium is stable. Theorem 1 requires the following Lemma 1, whose proof is omitted because it is similar to (and simpler than) the proof of Lemma 2 that is given in Appendix C.

**Lemma 1** *Consider Type I dynamics with threshold $\sigma$ and system (7) in arbitrary network $G = (V, E)$.*

(i) *If $\frac{1}{\deg(v)} \sum_{u \in N_v} B_u(0) > \sigma$ holds for all $v \in V$, then $\frac{1}{\deg(v)} \sum_{u \in N_v} B_u(t) > \sigma$ holds for all $v \in V$ and $t \geq 0$, and $\min_{v \in V} B_v(t)$ increases monotonically.*

(ii) *If $\frac{1}{\deg(v)} \sum_{u \in N_v} B_u(0) < \sigma$ holds for all $v \in V$, then $\frac{1}{\deg(v)} \sum_{u \in N_v} B_u(t) < \sigma$ holds for all $v \in V$ and $t \geq 0$, and $\max_{v \in V} B_v(t)$ decreases monotonically.*

**Theorem 1** *(a sufficient condition under which the defender or the attacker will occupy the entire network) Consider Type I dynamics with threshold $\sigma$ and arbitrary network $G = (V, E)$. If $\frac{1}{\deg(v)} \sum_{u \in N_v} B_u(0) > \sigma$ for all $v \in V$, $\lim_{t \to \infty} B_v(t) = 1$; if $\frac{1}{\deg(v)} \sum_{u \in N_v} B_u(0) < \sigma$ for all $v \in V$, $\lim_{t \to \infty} B_v(t) = 0$.*

**Proof** We only prove the first part because the second part can be proved analogously. According to Lemma 1, we know $\frac{1}{\deg(v)} \sum_{u \in N_v} B_u(t) > \sigma$ for all $t \geq 0$ and $v \in V$. This and Eq. (5) imply that $\theta_{v,RB}(t) = 1$ for all $t \geq 0$ and $v \in V$. Thus, system (7) becomes

$$\frac{dB_v(t)}{dt} = \theta_{v,RB}(t) - B_v(t) = 1 - B_v(t).$$

This leads to $B_v(t) = \exp(-t)B_v(0) + 1 - \exp(-t)$ and thus $\lim_{t \to \infty} B_v(t) = 1$. ∎

Theorem 1 holds for *arbitrary* networks, including the special case of complete graphs. Theorem 1 leads to the following insight (informally stated):

**Insight 1** *There is a quantitative relationship between the initial network security state and the combat-power function as indicated by the threshold $\sigma$ in Type I combat-power function. Specifically, when neither the defender nor the attacker is superior to its opponent, active cyber defense can automatically clean up a compromised network only when the defender has occupied more than a threshold $\sigma$ portions of the network (or nodes). This means that the defender may need to* manually *clean up some compromised nodes before using active cyber defense to* automatically *clean up the entire network.*

**Theorem 2** *(a sufficient condition under which neither the defender nor the attacker will occupy the entire network) Consider Type I dynamics with threshold $\sigma$ and arbitrary clustered network $G = (V, E)$. Let $B_v(0) = \alpha_k$ for every $v \in V_k$ and $\beta_k$ be the minimum node expansion as defined in Eq. (9). If $\alpha_k \beta_k > \sigma$, all nodes in $V_k$ will become* blue*; if $(1 - \alpha_k)\beta_k > 1 - \sigma$, all nodes in $V_k$ will become* red.

**Proof** If $\alpha_k \beta_k > \sigma$ for all nodes in $V_k$, then

$$\frac{1}{\deg(v)} \sum_{u \in N_v} B_u(0) \geq \frac{1}{\deg(v)} \alpha_k \cdot |N_v \cap V_k| \geq \alpha_k \beta_k > \sigma.$$

As in Theorem 1, we have $\lim_{t \to \infty} B_v(t) = 1$.

If $(1 - \alpha_k)\beta_k > 1 - \sigma$ for all nodes in $V_k$, then

$$1 - \frac{1}{\deg(v)} \sum_{u \in N_v} B_u(0) \geq \frac{\deg(v)}{\deg(v)} - \frac{|N_v \cap V_k| \cdot \alpha_k}{\deg(v)} - \frac{|N_v \setminus V_k|}{\deg(v)}$$

$$= \frac{|N_v \cap V_k|}{\deg(v)}(1 - \alpha_k) \geq (1 - \alpha_k)\beta_k > 1 - \sigma.$$

As in Theorem 1, we have $\lim_{t \to \infty} B_v(t) = 0$. ∎

Theorem 2, which applies to *arbitrary* networks with the cluster structure, leads to:

**Insight 2** *Suppose (i) neither the defender nor the attacker is superior to its opponent and (ii) the initial network security state does not satisfy the conditions of Theorem 1. Then, the network structure plays an important role. Specifically, in clustered networks, active cyber defense may only be able to automatically clean up some clusters, but not the entire network.*

Theorem 1 identifies two stable equilibria $B^* = [1, \ldots, 1]$ and $B^* = [0, \ldots, 0]$, while Theorem 2 gives a condition under which another kind of stable equilibria exist (i.e., different clusters in different color). Because the stability of equilibria gives a high-level description of Type I dynamics (e.g., under what condition the global network security state evolves toward a particular equilibrium), we need some general method/algorithm to evaluate the stability of equilibria. This is addressed by the following Theorem 3, whose proof is deferred to Appendix A. Before presenting the theorem, we recall that an equilibrium $B^*$ is *stable* if there exists a neighborhood of $B^*$ such that every trajectory $B(t)$ initially located in the neighborhood converges to $B^*$. We say $B^*$ is a stable equilibrium with exponential convergence if for each $B(t)$ in the neighborhood, there exist positive constants $\varrho > 0$ and $M > 0$ such that $\|B(t) - B^*\| \leq M e^{-\varrho t}$ for all $t \geq 0$.

**Theorem 3** *(method/algorithm for determining stability of equilibria and their emergence rates) Consider Type I dynamics with threshold $\sigma$ and arbitrary network $G = (V, E)$. Let $B^* = [B_v^*]_{v \in V}$ be an equilibrium and $\bar{B}^* = [1 - B_v^*]_{v \in V}$.*

(i) If the following holds for all $v \in V$

$$B_v^* = \begin{cases} 1 & \frac{1}{\deg(v)} \sum_{u \in N_v} B_u^* > \sigma \\ 0 & \frac{1}{\deg(v)} \sum_{u \in N_v} B_u^* < \sigma, \end{cases} \qquad (11)$$

both $B^*$ and $\bar{B}^*$ are asymptotically stable equilibria with *exponential* *convergence*.

(ii) If $B_v^* = \sigma$ for some $v \in V$, $B^*$ and $\bar{B}^*$ are unstable.

Recall that Theorem 1 says that the system has two equilibria: $[1, \ldots, 1]$ and $[0, \ldots, 0]$. Since both equilibria satisfy condition (11), Theorem 3 says that the two equilibria are asymptotically stable with *exponential* convergence.

### 4.2 Characterizing Type I Dynamics with Strategic Defender

Now we investigate Type I dynamics with *strategic* defender, where the initial probability that node $v$ is secure is proportional to its degree, namely $B_v(0) \propto \deg(v)$. We analyze it in the afore-reviewed generalized random graph model [40]. This means that our analytical result (Theorem 4 below) is not necessarily true for arbitrary networks. We compensate this slight restriction with valuable analytical results, including the quantification of the benefits when the attack-defense network structures are ER graphs and power-law graphs. The basic idea behind the proof of Theorem 4 is to show that under the given conditions, the event $\frac{1}{\deg(v)} \sum_{u \in N_v} B_u(0) > \sigma$ occurs almost surely. We accomplish this by using an asymptotical normal distribution, and by showing that the Lyapunov condition in the Central Limit Theorem and the Kolmogorov condition in the Strong Law of Large Numbers [41] are satisfied. The proof details are given in Appendix B.

**Theorem 4** (outcome of active cyber defense with strategic defender) *Let $G(n) = (V(n), E(n))$ be an instance of $n$-node random graph generated according to a given expected (in-)degree sequence $(d_1(n), \ldots, d_n(n))$. Given the degree-dependent probability $B_v(0)$, we determine $v$'s state according to $B_v(0)$ independent of anything else. Let $S = \{v : v \in V(n) \wedge B_v(0) = 1\}$ be the set of* blue *nodes in $G(n)$ at time $t = 0$, and*

$$\phi(n) = \frac{\sum_{v \in S} \deg(v)}{\sum_{u \in V(n)} \deg(u)},$$

*where $\deg(v)$ is the (in-)degree of $v \in V(n)$ in $G(n)$. Let*

$$s_{n,v}^2 = \sum_{u \in V(n)} B_u(0)^2 p_{vu}(n)(1 - p_{vu}(n)), \qquad (12)$$

$$q_{n,v} = \sum_{u \in V(n)} B_u(0)^3 p_{vu}(n)(1 - p_{vu}(n))\big[(1 - p_{vu}(n))^2 + p_{vu}(n)^2\big], \qquad (13)$$

$$w_{n,v}^2 = \sum_{u \in V(n)} p_{vu}(n)(1 - p_{vu}(n)), \qquad (14)$$

$$g_{n,v} = \sum_{u \in V(n)} p_{vu}(n)(1 - p_{vu}(n))\big[(1 - p_{vu}(n))^2 + p_{vu}(n)^2\big]. \qquad (15)$$

*Assume (i). $\lim_{n \to \infty} \sup_{v \in V(n)} q_{n,v}/s_{n,v}^3 = 0$; (ii). $\lim_{n \to \infty} \sup_{v \in V(n)} g_{n,v}/w_{n,v}^3 = 0$; (iii). $\lim_{n \to \infty} \sqrt{\ln(n)}/d_{\min}(n) = 0$; (iv). $\lim_{n \to \infty} (\sum_{v \in V(n)} g_{n,v})/(\sum_{v \in v(n)} w_{n,v}^2)^{3/2} = 0$; (v). $\lim_{n \to \infty} (\sum_{v \in V(n)} q_{n,v})/(\sum_{v \in v(n)} s_{n,v}^2)^{3/2} = 0$; (vi). $\lim_{n \to \infty} \sum_{v \in V(n)} \frac{1}{d_v^2} = 0$. If $\underline{\lim}_{n \to \infty} \phi(n) > \sigma$ holds almost surely, then $\lim_{n \to \infty} \lim_{t \to \infty} B_v(t) = 1$ holds for all $v \in V(n)$ almost surely, namely $\lim_{n \to \infty} \mathsf{P}(\lim_{t \to \infty} B_v(t) = 1) = 1$; if $\overline{\lim}_{n \to \infty} \phi(n) < \sigma$ holds almost surely, then $\lim_{n \to \infty} \lim_{t \to \infty} B_v(t) = 0$ holds for all $v \in V(n)$ almost surely, namely $\lim_{n \to \infty} \mathsf{P}(\lim_{t \to \infty} R_v(t) = 1) = 1$.*

Note that Theorem 4 holds for generalized random graphs (rather than arbitrary networks), which however are not necessarily dense. To see this, we observe that a sufficient condition for assumption (v) is $d_{\min} \gg \sqrt{n}$ because

$$\sum_{v \in V(n)} \frac{1}{d_v^2(n)} \leq \frac{n}{d_{\min}^2(n)}.$$

A necessary condition for assumption (v) is $\langle d_v^2(n) \rangle \gg n$, where $\langle d_v^2(n) \rangle = \frac{1}{n} \sum_{v \in V(n)} d_v^2(n)$, because

$$\sum_{v \in V(n)} \frac{1}{d_v^2(n)} \geq \frac{n}{\frac{1}{n} \sum_{v \in V(n)} d_v^2(n)} = \frac{n}{\langle d_v^2(n) \rangle}.$$

These conditions do *not* imply that the graphs are dense. For example, the two conditions are satisfied by $d_v(n) = O(\sqrt{n} \log(n))$ for all $v \in V(n)$, which however implies that density of the graph converges to zero as $n \to \infty$.

Theorem 4 corresponds to the case of strategic defender with $B_v(0) \propto \deg(v)$, and can be adapted to the case of strategic attacker with $R_v(0) \propto \deg(v)$. In what follows we discuss the implications of Theorem 4 in these two cases separately, and then compare them to draw deeper/quantitative insights with respect to ER and power-law graphs.

**Characterizing the qualitative benefit to strategic defender with $B_v(0) \propto \deg(v)$**

Since $B_v(0) \propto \deg(v)$, we have $B_v(0) = C_1 \frac{\deg(v)}{\sum_{u \in V(n)} \deg(u)}$ for some constant $C_1 > 0$. Then, the expected number of initial blue nodes is

$$\sum_{v \in V(n)} B_v(0) = C_1 \sum_{v \in V(n)} \frac{\deg(v)}{\sum_{u \in V(n)} \deg(u)} = C_1.$$

Define

$$\alpha_{threshold} = \frac{\sigma}{n} \frac{[\sum_{u \in V(n)} \deg(u)]^2}{\sum_{v \in V(n)} \deg(v)^2}, \tag{16}$$

where $\deg(v)$ is the (in-)degree of node $v \in V(n)$. With respect to random set $S$ of blue nodes at time $t = 0$, we define random variable $\chi_v(S)$:

$$\chi_v(S) = \begin{cases} 1 & v \in S \\ 0 & v \notin S. \end{cases}$$

Since

$$\phi(n) = \frac{\sum_{u \in S} \deg(u)}{\sum_{v \in V(n)} \deg(v)} = \frac{\sum_{u \in V(n)} \deg(u) \chi_u(S)}{\sum_{v \in V(n)} \deg(v)}$$
$$\approx \frac{\sum_{u \in V(n)} \deg(u) B_v(0)}{\sum_{v \in V(n)} \deg(v)} > \sigma, \tag{17}$$

Theorem 4 implies the following: if $\frac{|S|}{n} > \alpha_{threshold}$, then $\lim_{t \to \infty} B_v(t) = 1$ for $v \in V(n)$; if $\frac{|S|}{n} < \alpha_{threshold}$, then $\lim_{t \to \infty} B_v(t) = 0$ for $v \in V(n)$. Since $\frac{[\sum_{u \in V(n)} \deg(u)]^2}{\sum_{v \in V(n)} \deg(v)^2} \leq n$, we have $\alpha_{threshold} \leq \sigma$. This means that a strategic defender can use active cyber defense to automatically clean up the entire network *even if* the defender initially occupies less than $\sigma$, but more than $\alpha_{threshold}$ ($\leq \sigma$), portions of the network. This leads to:

**Insight 3** *If the large-degree nodes are appropriately better protected by the strategic defender, the strategic defender can use active cyber defense to automatically clean up the network* even if *it only occupies $\alpha_{threshold}$ ($\leq \sigma$) portions of the network.*

**Characterizing the qualitative benefit to strategic attacker with $R_v(0) \propto \deg(v)$**

In this case, we have $R_v(0) = C_2 \frac{\deg(v)}{\sum_{u \in V} \deg(u)}$ for some constant $C_2 > 0$. According to Eq. (8), $f_{BR}(\cdot)$ is discontinuous at $1 - \sigma$. The red-node initial occupation threshold is $\frac{1-\sigma}{n} \frac{[\sum_{u \in V} \deg(u)]^2}{\sum_{v \in V} \deg(v)^2}$. Thus, the blue-node initial occupation threshold is

$$\beta_{threshold} = 1 - \frac{1-\sigma}{n} \frac{[\sum_{v \in V} \deg(v)]^2}{\sum_{v \in V} \deg(v)^2}. \tag{18}$$

If $\frac{|S|}{n} > \beta_{threshold}$, $\lim_{t \to \infty} B_v(t) = 1$; if $\frac{|S|}{n} < \beta_{threshold}$, $\lim_{t \to \infty} B_v(t) = 0$. Since $\beta_{threshold} \geq \sigma$, this leads to:

**Insight 4** *If the large-degree nodes are compromised by the strategic attacker, the defender can use active defense to clean up the network only after the defender occupies $\beta_{threshold}$ ($\geq \sigma$) portions of the network.*

**Characterizing the quantitative benefit to strategic defender in ER graphs**

For ER graphs with edge probability $p$, the degree distribution follows a binomial distribution $B(n, p)$:

$$P(\deg(v) = k) = \binom{n}{p} p^k (n-p)^k, \ k = 0, 1, \ldots.$$

In the above we showed

$$\alpha_{threshold} = \sigma \frac{p}{p + p(1-p)/n}, \quad \beta_{threshold} = 1 - (1-\sigma) \frac{p}{p + p(1-p)/n}.$$

As $n \to \infty$, both $\alpha_{threshold}$ and $\beta_{threshold}$ converge to the threshold $\sigma$. More specifically, $\beta_{threshold} - \alpha_{threshold} = 1 - p/[p + p(1-p)/n]$ converges to 0, while $\frac{\beta_{threshold}}{\alpha_{threshold}} = 1 + \frac{1-p}{\sigma n}$ converges to 1. This leads to:

**Insight 5** *For large ER graphs, the benefit to strategic defender/attacker is* not *significant because the node degrees are relatively homogeneous. (This is reminiscent of, and in parallel to, the* connectivity-based *robustness of ER networks, namely that ER networks are resilient against strategic deletion of large-degree nodes [12]. Note however that in our model, the attacker aims to compromise nodes but does* not *delete any nodes.)*

**Characterizing the quantitative benefit to strategic defender in power-law graphs**

Consider power-law graphs with exponent $\gamma$. Let $C = \int_{d_{\min}(n)}^{d_{\max}(n)} k^{-\gamma} dk = \frac{d_{\max}(n)^{1-\gamma} - d_{\min}(n)^{1-\gamma}}{1-\gamma}$. By replacing the sum with integral in Eq. (16), we can define

$$\begin{aligned}\alpha_{threshold} &= \sigma \left( \frac{n}{C} \int_{d_{\min}(n)}^{d_{\max}(n)} k^{1-\gamma} dk \right)^2 \Big/ \left( \frac{n}{C} \int_{d_{\min}(n)}^{d_{\max}(n)} k^{2-\gamma} dk \right) \\ &= \frac{\sigma}{n} \left( \frac{n^2 (d_{\max}(n)^{2-\gamma} - d_{\min}(n)^{2-\gamma})^2/(2-\gamma)^2}{(d_{\max}(n)^{1-\gamma} - d_{\min}(n)^{1-\gamma})^2/(1-\gamma)^2} \right) \Big/ \left( \frac{n(d_{\max}(n)^{3-\gamma} - d_{\min}(n)^{3-\gamma})/(3-\gamma)}{(d_{\max}(n)^{1-\gamma} - d_{\min}(n)^{1-\gamma})/(1-\gamma)} \right).\end{aligned}$$

This leads to four cases: $\gamma \notin \{1, 2, 3\}$, $\gamma = 1$, $\gamma = 2$, $\gamma = 3$. Let $z = d_{\max}(n)/d_{\min}(n)$. For $\gamma \notin \{1, 2, 3\}$, one can show

$$\begin{aligned}&\frac{(d_{\max}(n)^{2-\gamma} - d_{\min}(n)^{2-\gamma})^2/(2-\gamma)^2}{(d_{\max}(n)^{1-\gamma} - d_{\min}(n)^{1-\gamma})^2/(1-\gamma)^2} \Big/ \frac{(d_{\max}(n)^{3-\gamma} - d_{\min}(n)^{3-\gamma})/(3-\gamma)}{(d_{\max}(n)^{1-\gamma} - d_{\min}(n)^{1-\gamma})/(1-\gamma)} \\ &= \frac{(z^{2-\gamma} - 1)^2}{(z^{1-\gamma} - 1)(z^{3-\gamma} - 1)} \frac{(3-\gamma)(1-\gamma)}{(2-\gamma)^2}.\end{aligned}$$

For $\gamma = 1, 2, 3$, we can reason in a similar fashion. As a result, we can define

$$h(z, \gamma) = \begin{cases} \frac{(z^{2-\gamma}-1)^2}{(z^{1-\gamma}-1)(z^{3-\gamma}-1)} \cdot \frac{(3-\gamma)(1-\gamma)}{(2-\gamma)^2} & \gamma \neq 1, 2, 3 \\ 2\frac{z-1}{z+1}\frac{1}{\ln(z)} & \gamma = 1 \\ \frac{z(\ln(z))^2}{(z-1)^2} & \gamma = 2 \\ 2\frac{z-1}{z+1}\frac{1}{\ln(z)} & \gamma = 3. \end{cases}$$

If the defender is strategic, a sufficient condition for $\lim_{t \to \infty} B_v(t) = 1$ is $\frac{|S|}{n} > \alpha_{threshold} = \sigma \cdot h(z, \gamma)$; if the attacker is strategic, a sufficient condition for $\lim_{t \to \infty} B_v(t) = 1$ is $\frac{|S|}{n} > \beta_{threshold} = 1 - (1-\sigma)h(z, \gamma)$. Therefore, we have

$$\beta_{threshold} - \alpha_{threshold} = 1 - h(z, \gamma), \qquad (19)$$

$$\frac{\beta_{threshold}}{\alpha_{threshold}} = \frac{1 - (1-\sigma)h(z, \gamma)}{\sigma \cdot h(z, \gamma)} = 1 + \frac{1 - h(z, \gamma)}{\sigma \cdot h(z, \gamma)}. \qquad (20)$$

Eqs. (19) and (20) reach maximum at $\gamma = 2$. This leads to:

**Insight 6** *For power-law graphs, the benefit to strategic defender/attacker is significant. (This is also reminiscent of, and in parallel to, the* connectivity-based *robustness of power-law networks, namely that power-law networks are easily disrupted by strategic deletion of large-degree nodes [12]. Again, in our model the attacker aims to compromise nodes but does not delete any nodes.) Moreover, the benefit to strategic defender is maximized for the sub-class of power-law networks with exponent $\gamma = 2$.*

## 5 Characterizing Type II Active Cyber Defense Dynamics

Type II dynamics is similar to Type I dynamics, except the following: Type-I combat-power function is discontinuous near the threshold $\sigma$, whereas Type II combat-power function is continuous and differentiable near the threshold $\tau$. For the case of *non-strategic* defender with node-independent $B_v(0)$, we obtain the following Theorems 5-7, which are in parallel to Theorems 1-3, respectively. Theorem 5 requires the following Lemma 2, whose proof is given in Appendix C.

**Lemma 2** *Consider Type II dynamics with threshold $\tau$ and system (7) in arbitrary network $G = (V, E)$.*

(i) *If $\frac{1}{\deg(v)} \sum_{u \in N_v} B_u(0) > \tau$ holds for all $v \in V$, then $\frac{1}{\deg(v)} \sum_{u \in N_v} B_u(t) > \tau$ holds for all $v \in V$ and $t \geq 0$, and $\min_{v \in V} B_v(t)$ increases monotonically.*

(ii) *If $\frac{1}{\deg(v)} \sum_{u \in N_v} B_u(0) < \tau$ holds for all $v \in V$, then $\frac{1}{\deg(v)} \sum_{u \in N_v} B_u(t) < \tau$ holds for all $v \in V$ and $t \geq 0$, and $\max_{v \in V} B_v(t)$ decreases monotonically.*

Proof of the following Theorem 5, which holds for *arbitrary* networks, is given in Appendix D.

**Theorem 5** *(a sufficient condition under which the defender or the attacker will occupy the entire network) Consider Type II dynamics with threshold $\tau$ and arbitrary network $G = (V, E)$. If $\frac{1}{\deg(v)} \sum_{u \in N_v} B_u(0) > \tau$ for all $v \in V$, $\lim_{t \to \infty} B_v(t) = 1$ for all $v \in V$; if $\frac{1}{\deg(v)} \sum_{u \in N_v} B_u(0) < \tau$ for all $v \in V$, $\lim_{t \to \infty} B_v(t) = 0$ for all $v \in V$.*

**Theorem 6** *(a sufficient condition under which neither the defender nor the attacker will occupy the entire network) Consider Type II dynamics with threshold $\tau$ and* arbitrary *network $G = (V, E)$ with the cluster structure. Let $B_v(0) = \alpha_k$ for every $v \in V_k$ and $\beta_k$ be the minimum node expansion as defined in Eq. (9). If $\alpha_k \beta_k > \tau$, all nodes in $V_k$ will become* blue*; if $(1-\alpha_k)\beta_k > 1-\tau$, all nodes in $V_k$ will become* red.

Proof of Theorem 6 is similar to proof of Theorem 2. Proof of the following Theorem 7 is given in Appendix E. Both theorems hold for *arbitrary* networks.

**Theorem 7** *(method/algorithm for determining stability of equilibria) Consider Type II dynamics with threshold $\tau$ and* arbitrary *network $G = (V, E)$. Let $B^* = [B_v^*]_{v \in V}$ be an equilibrium and $\bar{B}^* = [1 - B_v^*]_{v \in V}$.*

(i) *Equilibria $B^* = [1, \ldots, 1]$ and $B^* = [0, \ldots, 0]$ are asymptotically stable with* exponential *convergence.*

(ii) *If $B_v^* = \tau$ for some $v \in V$, $B^*$ and $\bar{B}^*$ are unstable.*

For the case of *strategic* defender with $B_v(0) \propto \deg(v)$, we can obtain a result for generalized random graphs in parallel to Theorem 4 via a similar proof. We omit the lengthy details. In summary, we have:

**Insight 7** *The preceding* **Insights** *1-6 are equally applicable to Type II dynamics.*

## 6 Characterizing Types III-IV Active Cyber Defense Dynamics

Types III-IV combat-power functions represent that the defender (attacker) is superior to, or more advanced than, its opponent. Due to the lack of threshold in the computer-power functions, an immediate consequence is that there is no difference between the case of non-strategic defender and the case of strategic defender. Further consequences due to the lack of threshold are characterized as follows.

**Theorem 8** *(characterizing Type III dynamics) Consider Type III dynamics in* arbitrary *network $G = (V, E)$.*

(i) *If $B_v(0) > 0$ for all $v \in V$, then $\lim_{t \to \infty} B_v(t) = 1$ for all $v \in V$.*

(ii) *Equilibrium $B_v(0) = [1, \ldots, 1]$ is asymptotically stable with* exponential *convergence.*

(iii) *Equilibrium $B_v(0) = [0, \ldots, 0]$ is unstable.*

Part (i) of Theorem 8 can be proved as in the first half of Theorem 5. Parts (ii) can be proved in a fashion similar to Part (i) of Theorem 7. Parts (iii) can be proved in a fashion similar to Part (ii) of Theorem 7. Since Type IV dynamics is dual to Type III dynamics, from Theorem 8 we obtain:

**Theorem 9** *(characterizing Type IV dynamics) Consider Type IV dynamics in* arbitrary *network $G = (V, E)$.*

(i) *If $B_v(0) < 1$ for all $v \in V$, then $\lim_{t \to \infty} B_v(t) = 0$ for all $v \in V$.*

(ii) *Equilibrium $B_v(0) = [0, \ldots, 0]$ is asymptotically stable with* exponential *convergence.*

(iii) *Equilibrium $B_v(0) = [1, \ldots, 1]$ is unstable.*

Theorems 8-9, which hold for *arbitrary* networks, lead to:

**Insight 8** *If the defender is superior to the attacker in terms of cyber combat power, the defender can always use active defense to automatically clean up the entire network as long as there are a few computers that are not compromised. In the extreme case where the attacker compromised the entire network, the defender only needs to manually clean up a few computers before launching active defense to automatically clean up the entire network. These suggest that cyber combat superiority can serve as an effective deterrence.*

# 7 Advantage of Active Cyber Defense over Reactive Cyber Defense

Current cyber defense is mainly reactive, where the defender runs "anti-virus software"-like tools on each computer to scan and cure infections, which are caused by attacks/malwares that penetrated the perimeter defense (e.g., Firewalls). Reactive cyber defense inevitably causes an asymmetry that is advantageous to the attacker because the attack effect is automatically amplified by the network (a kind of "network effect"). Specifically, reactive cyber defense may be modeled using the well-known SIS (Susceptible-Infectious-Susceptible) model, while accommodating arbitrary attack-defense network topologies. A sufficient condition for the spreading to die out is [16]:

$$\lambda_{1,\mathsf{A}} < \frac{cure\_capability}{spreading\_capability},$$

where $\lambda_{1,\mathsf{A}}$ is the largest eigenvalue of the adjacency matrix corresponding to the attack-defense defense structure and is in a sense the average node degree or connectivity [42], $cure\_capability$ abstracts the defender's reactive defense power (i.e., the probability that a compromised node becomes a susceptible node at a single time step), and $spreading\_capability$ abstracts the attacker's attack power (i.e., the probability that a compromised node successfully attacks a susceptible neighboring node at a single time step). This means that the attacker always benefits from rich connectivity because the attack effect is amplified by $\lambda_{1,\mathsf{A}}$, which explains why the asymmetry phenomenon is advantageous to the attacker [43, 44, 45].

On the other hand, Sections 4-6 show that the asymmetry disappears with active cyber defense because $\lambda_{1,\mathsf{A}}$ (or its like) does not play a role in the analytical results. This justifies one usefulness of the model-based characterization studies. In summary, we have:

**Insight 9** *Active cyber defense eliminates the attack amplification phenomenon, namely the asymmetry between cyber attack and reactive cyber defense.*

# 8 Validating the Dynamic System Model via Simulation

The above characterizations of active cyber defense dynamics are based on the Dynamic System model, which is the mean-field approximation of the native Markov process model. Therefore, we need to show whether or not the analytical results derived from the Dynamic System model are inherent to the Markov process model.

## 8.1 Validation Methodology

Our validation methodology is centered on examining the *dynamics accuracy* and the *threshold accuracy* of the Dynamic System model. For examining the dynamic accuracy, we compare the mean blue occupation probability in the Dynamic System model, namely $\langle B_v(t) \rangle = \frac{1}{|V|} \sum_{v \in V} B_v(t)$, and the simulation-based mean fraction of blue nodes in the Markov process model, namely $\langle \xi_v(t) \rangle = \frac{1}{|V|} \sum_{v \in V} \xi_v(t)$. If $\langle \xi_v(t) \rangle$ and $\langle B_v(t) \rangle$ exhibit a similar, if not exactly the same, dynamic behavior, we conclude that the analytical results derived from the Dynamic System model are inherent to the Markov process model. (i) Our simulation of the Markov process model is based on Eq. (1), namely

$$\mathsf{P}\{\xi_v(t+\Delta t) = 1 | \xi_v(t),\ v \in N\} = \begin{cases} \Delta t \cdot \tilde{\theta}_{v,RB}(t) & \xi_v(t) = 0 \\ 1 - \Delta t \cdot \tilde{\theta}_{v,BR}(t) & \xi_v(t) = 1 \end{cases}$$

where the random rate $\tilde{\theta}_{v,RB}$ is replaced with its mean $\theta_{v,RB}$ as specified in Eq. (5). Simulation results are based on the average of 50 simulation runs. (ii) Our numerical calculation in the Dynamic System model is based on Eq. (7),

namely

$$B_v(t + \Delta t) = B_v(t) + [\theta_{v,RB}(t) - B_v(t)]\Delta t.$$

In both cases, we set $\Delta t = 0.01$.

For examining the threshold accuracy, we study whether or not the threshold $\sigma$ in the Dynamic System model is faithful to the threshold $\sigma_{markov}$ in the Markov process model. In order to to compute $\sigma_{markov}$, we use the following numerical method. Since the convergence of $\langle \xi_v(t) \rangle$ is probabilistic in a very small interval that contains $\sigma$, we define $\sigma_{markov}$ as the median value in that interval. Specifically, let $a_1$ be the smallest value such that an initial blue occupation greater than $a_1$ will cause all nodes to become blue in all 50 runs. Let $b_1$ be the largest value so that an initial blue occupation smaller than $b_1$ will cause all nodes to become red in all 50 runs. We set $\sigma_{markov} = \frac{a_1+b_1}{2}$.

In our simulation, we use two kinds of graphs:

- ER random graph: It has $n = 2,000$ nodes and independent link probability $p = 0.02$.

- Power-law random graph: It has $n = 2,000$, exponent $\gamma = 2.5$, minimum node degree 2, and maximum node degree 120.

## 8.2 Dynamics Accuracy of the Dynamic System Model

**Overall dynamics accuracy**

First, let us consider Type I dynamics and non-strategic defender with node-independent identical initial occupation probability $B_v(0)$. Figure 4 confirms that Theorem 1, which was proven in the Dynamic System model, is indeed inherent to the Markov process model. Specifically, in the Dynamic System model, the $\langle B_v(t) \rangle$'s corresponding to $B_v(0) = 0.4 > \sigma = 1/3$ all converge to 1, and the $\langle B_v(t) \rangle$'s corresponding to $B_v(0) = 0.2 < \sigma = 1/3$ all converge to 0. In the Markov process model, the $\langle \xi_v(t) \rangle$'s corresponding to $\mathsf{P}\{\xi_v(0) = 1\} = 0.4$ all converge to 1, and the $\langle \xi_v(t) \rangle$'s corresponding to $\mathsf{P}\{\xi_v(0) = 1\} = 0.2$ all converge to 0. Therefore, the dynamic behavior indicated by Theorem 1 is also exhibited by the Markov process model.

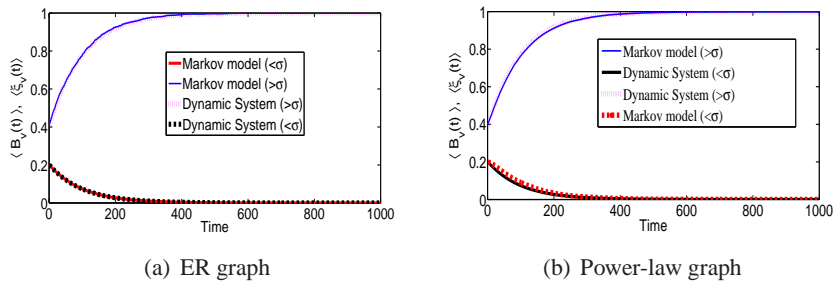

(a) ER graph      (b) Power-law graph

Figure 4: $\langle B_v(t) \rangle$ vs. $\langle \xi_v(t) \rangle$ in Type I dynamics with $\sigma = 1/3$ and non-strategic defender.

Second, let us look at Type I dynamics and strategic defender with $B_v(0) \propto \deg(v)$. Define $\eta = \frac{\sum_{u \in S} \deg(u)}{\sum_{v \in V} \deg(v)}$, where $S$ is the set of blue nodes at time $t = 0$. Inequality (17) indicates that if $\eta > \sigma$, all nodes will become blue; if $\eta < \sigma$, all nodes will become red. In our simulation, we set $\sigma = 0.5$. Figure 5(a) shows that in the ER graph, both $\langle B_v(t) \rangle$ in the Dynamic System model and $\langle \xi_v(t) \rangle$ in the Markov process model converge to 1 when $\eta = 0.52 > \sigma = 0.5$, and converge to 0 when $\eta = 0.45 < \sigma = 0.5$. Figure 5(b) shows that in the power-law network, both $\langle B_v(t) \rangle$ and $\langle \xi_v(t) \rangle$ converge to 1 when $\eta = 0.45$ ($< \sigma = 0.5$), and converge to 0 when $\eta = 0.35$ (far smaller than $\sigma = 0.5$). These confirm the phenomenon that is implied by Theorem 4, namely that the effect of strategic defense is not significant in ER networks but significant in power-law networks. In any case, the simulation results demonstrate that the phenomenon exhibited by the Dynamic System model is inherent to the Markov process model.

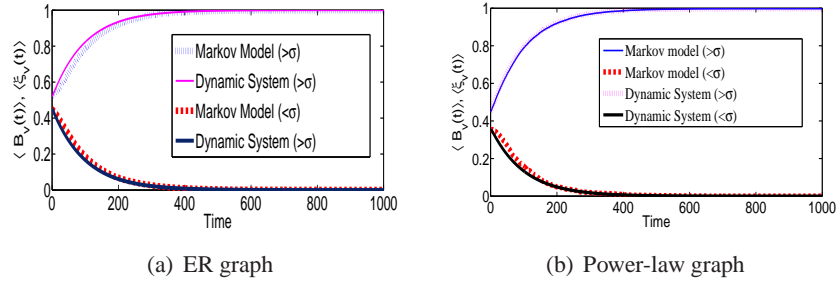

(a) ER graph

(b) Power-law graph

Figure 5: $\langle B_v(t) \rangle$ vs. $\langle \xi_v(t) \rangle$ in Type I dynamics with $\sigma = 1/2$ and strategic defender.

Third, let us look at Types II-IV dynamics and non-strategic defender with node-independent identical initial occupation probability $B_v(0)$. Consider Type II combat-power function with $\tau = 0.5$, $f_{RB}(x) = 2x^2$ for $x \in [0, 0.5]$, and $f_{RB}(x) = -2x^2 + 4x - 1$ for $x \in [0.5, 1]$. For the Dynamic System model, Figures 6(a)-6(b) show $B_v(0) = 0.4 < \tau = 0.5$ implies that all nodes will become red, and $B_v(0) = 0.6 > \tau = 0.5$ implies that all nodes will become blue. In the Markov process model, the same phenomenon is exhibited with the same initial condition $\mathsf{P}\{\xi_v = 1\} = B_v(0)$. This validates Theorem 5. For Type III combat-power function with $f_{RB}(x) = x^{1/2}$, Figures 6(c)-6(d) demonstrate that $\langle B_v(t) \rangle$ corresponding to $B_v(0) = 0.02$ converges to 1 in the Dynamic System model. The same phenomenon is exhibited in the Markov process model. This validates Theorem 8. For Type IV combat-power function $f_{RB}(x) = x^2$, Figures 6(e)-6(f) validate that $\langle B_v(t) \rangle$ corresponding to $B_v(0) = 0.98$ converges to 0. The same phenomenon is exhibited in the Markov process model. This confirms that the dynamic behavior indicated by Theorem 9 is also exhibited by the Markov process model.

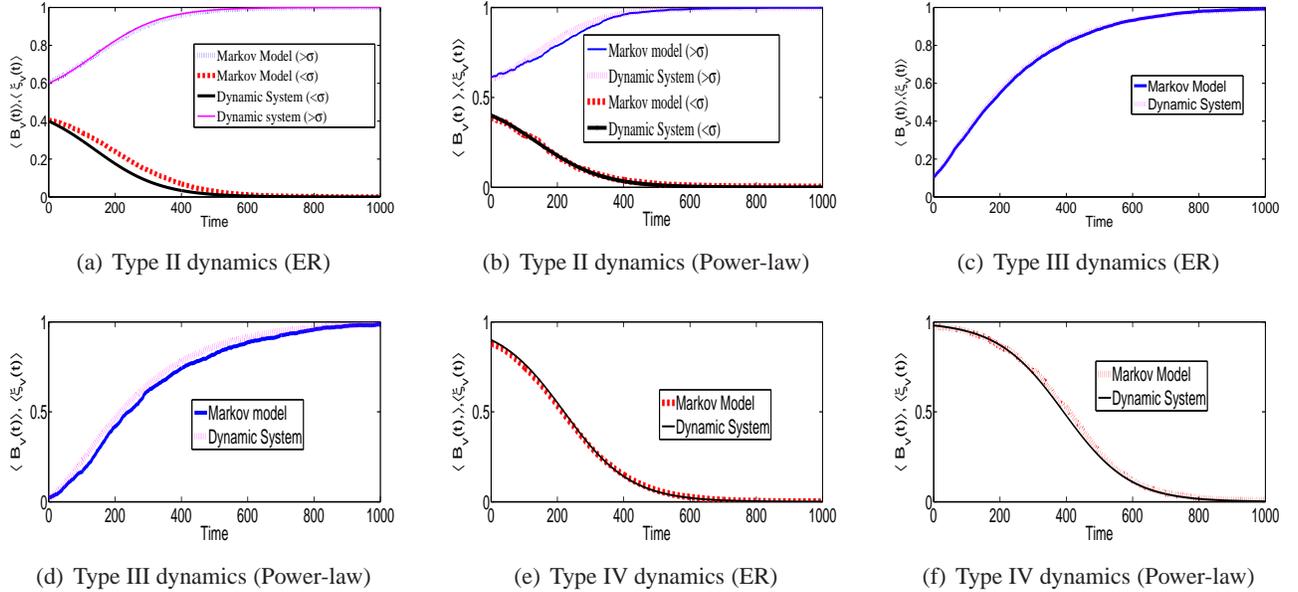

(a) Type II dynamics (ER)

(b) Type II dynamics (Power-law)

(c) Type III dynamics (ER)

(d) Type III dynamics (Power-law)

(e) Type IV dynamics (ER)

(f) Type IV dynamics (Power-law)

Figure 6: $\langle \xi_v(t) \rangle$ vs. $\langle B_v(t) \rangle$ in Types II-IV dynamics with non-strategic defender.

Fourth, for power-law networks and strategic defender with $B_v(0) \propto \deg(d)$, we derived the sufficient condition

$\frac{|S|}{n} > \sigma \cdot h(z,\gamma)$ for $\lim_{t\to\infty} B_v(t) = 1$, meaning that in order for the defender to use active cyber defense to automatically clean up the network, the defender needs to occupy more than $\sigma \cdot h(z,\gamma)$ fraction of the nodes, which is minimum when $h(z,\gamma)$ is minimum. As shown in Figure 7(a), for fixed $z$, $h(z,\gamma)$ is minimum at $\gamma=2$, which corresponds to the sub-class of power-law networks that *maximize the benefit to the strategic defender*. Figure 7(b) plots the simulation results in the Markov process model. We observe that $\sigma_{markov}$ is minimum at $\gamma = 2$ in the Markov model as well. These further confirm that the particular conclusion drawn in the Dynamic model — the benefit to the strategic defender is maximized for power-law graphs with exponent $\gamma = 2$ — is also inherent to the Markov process model.

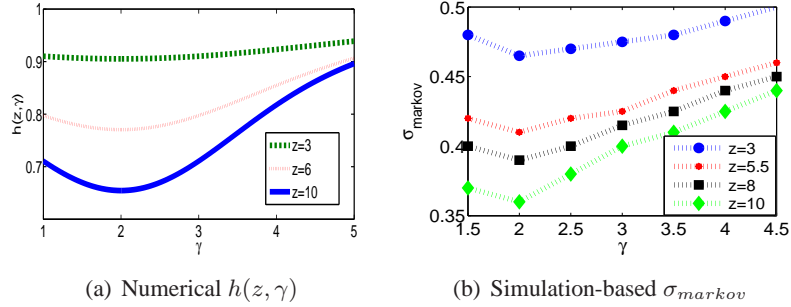

(a) Numerical $h(z,\gamma)$  (b) Simulation-based $\sigma_{markov}$

Figure 7: Power-law networks with exponent $\gamma = 2$ maximizes the benefit to strategic defenders

**Dynamics inaccuracy: cause and characteristics**

In the above, our simulation results show, from the perspective of system state dynamics, that the Dynamic System model offers overall accurate approximation to the Markov process model. Still, Figures 4-6 visually exhibit the following phenomenon: the Dynamic System model sometimes underestimates and sometimes overestimates the dynamics simulated from the Markov process model. What is the cause of this phenomenon? To answer this question, we observe that the master equation Eq. (3) can be rewritten as:

$$\frac{d}{dt}\tilde{B}_v(t) = \tilde{\theta}_{v,RB}(t) - \tilde{B}_v(t) \tag{21}$$

where

$$\tilde{\theta}_{v,RB}(t) = \mathsf{E}\left(f_{RB}\left(\frac{1}{\deg(v)}\sum_{u \in N_v} \xi_v(t)\right)\right)$$

and $\tilde{\theta}_{v,RB}(t) = 1 - \tilde{\theta}_{v,BR}(t)$ so as to be consistent with Eq. (6). It can be seen that if $f_{RB}(\cdot)$ is convex, then

$$\begin{aligned}\tilde{\theta}_{v,RB}(t) &= \mathsf{E}\left(f_{RB}\left(\frac{1}{\deg(v)}\sum_{u \in N_v} \xi_v(t)\right)\right) \\ &\geq f_{RB}\left(\mathsf{E}\left(\frac{1}{\deg(v)}\sum_{u \in N_v} \xi_v(t)\right)\right) = \theta_{v,RB}(t).\end{aligned}$$

Analogously, if $f_{RB}(\cdot)$ is concave, then $\tilde{\theta}_{v,RB}(t) \leq \theta_{v,RB}(t)$. As a result, the above phenomenon can be explained as follows: For Types I-II combat-power functions, the dynamics in the Dynamic System model underestimates the dynamics in the native Markov process model when $\frac{1}{\deg(v)}\sum_{u \in N_v} \xi_v(t)$ is below the threshold in the combat-power function, and overestimates the dynamics in the Markov process model when $\frac{1}{\deg(v)}\sum_{u \in N_v} \xi_v(t)$ is above the threshold (see Figures 4-5 and Figures 6(a)-6(b)). For Type III combat-power functions, which can be regarded as concave over the region $[0,1]$, the dynamics in the Dynamic System model overestimates the dynamics in the Markov process model (see Figures 6(c)-6(d)). Analogously, for Type IV combat-power functions, the dynamics of the Dynamic System model underestimates the dynamics in the Markov process model (see Figures 6(e)-6(f)).

Having explained the cause of the slight dynamic inaccuracy, we want to establish some deeper understanding of the inaccuracy. In particular, we want to know how the inaccuracy may be dependent upon the average node degree. For this purpose, we consider the following notion of *relative error* between the Dynamic System model and the Markov process model:

$$RE = \frac{\int_0^T [\tilde{B}_v(t) - B_v(t)]^2 dt}{\int_0^T \tilde{B}_v^2(t) dt},$$

where $\tilde{B}_v(t)$ is the probability that node $v$ is blue in the Markov process model, and $B_v(t)$ is the dynamic system state. To investigate the impact of average node degree, we fix the variance of the node degrees, denoted by dvar. Consider in the generalized random graph model with a given expected degree sequence that follows the power-law distribution. By fixing the variance dvar and the ratio r between the minimum and maximum expected degrees as $d_{\max} = r * d_{\min}$, we derive $d_{\min}$ with respect to the varying power-law exponent $\gamma$ from 1 to 6, as follows:

$$d_{\min} = \sqrt{\frac{\text{dvar}}{\frac{1-\gamma}{3-\gamma}\frac{r^{3-\gamma}-1}{r^{1-\gamma}-1} - \frac{1-\gamma}{(2-\gamma)^2}\frac{r^{2-\gamma}-1}{(r^{1-\gamma}-1)^2}}}.$$

With r $= 20$ and dvar $= 400$, we obtain a series of generalized random graphs of 2,000 nodes. Although we cannot precisely fix the variance, the actual standard deviation of degrees for different $\gamma$'s is quite stable: $20.47 \pm 0.48$. We run the Markov process model and the Dynamic System model on the random graphs to calculate the relative errors. We find, as shown in Figure 8, that the relative errors decrease with the average node degree.

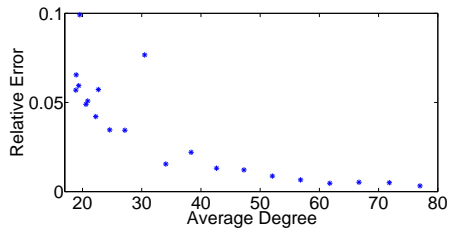

Figure 8: Relative error vs. average node degree

## 8.3 Threshold Accuracy of the Dynamic System Model

Now we examine the accuracy of the Dynamic System model from a different perspective: *threshold accuracy*. That is, we examine the accuracy of the threshold $\sigma$ derived from the Dynamic System model with respect to the threshold $\sigma_{markov}$, which is numerically derived from the Markov process model. For the special case of Type III-IV combat-power functions, which have no threshold, we observe the following: For Type III combat-power functions, if $\tilde{B}_v(0) > 0$ for some nodes that can reach all other nodes, then $\lim_{t \to \infty} \tilde{B}_v(t) = 1$ for all $v \in V$; For Type IV combat-power functions, if $\tilde{R}_v(0) > 0$ for some nodes that can reach all other nodes, then $\lim_{t \to \infty} \tilde{B}_v(t) = 0$ for all

$v \in V$. To see this, we note that in the case of Type III combat-power functions, the following holds:

$$\tilde{\theta}_{v,RB}(t) = \mathsf{E}\left[f\left(\frac{1}{\deg(v)}\sum_{u \in N_v}\xi_u(t)\right)\right]$$

$$\geq \mathsf{E}\left[\frac{1}{\deg(v)}\sum_{u \in N_v}\xi_u(t)\right] = \left[\frac{1}{\deg(v)}\sum_{u \in N_v}\tilde{B}_u(t)\right].$$

The case of Type IV combat-power functions can be treated analogously. However, the situation for Types I-II combat-power functions is very different as we elaborate below.

### Threshold (in)accuracy for Types I-II combat-power functions: cause and characteristics

We illustrate the following *threshold-drifting* phenomenon with the specific $f_{RB}(\cdot)$ in Type I dynamics for example. Figures 9(a)-9(b) plot $\sigma_{markov}$ and $\sigma$ in the case of non-strategic defender with node-independent identical probability $B_v(0)$. Figures 9(c)-9(d) plot $\sigma_{markov}$ and $\sigma$ in the case of strategic defender with $B_v(0) \propto \deg(v)$. We observe that Figure 9(d) exhibits a pattern that is different from the others, which we cannot explain at the moment but we plan to investigate in the future. In all other cases, we observe the following: if $\sigma < 0.5$, then $\sigma_{markov} < \sigma$; if $\sigma > 0.5$, then $\sigma_{markov} > \sigma$. We call this *threshold-drifting* phenomenon, which indicates that the threshold $\sigma$ in the Dynamic System model may deviate from the threshold $\sigma_{markov}$ in the Markov process model.

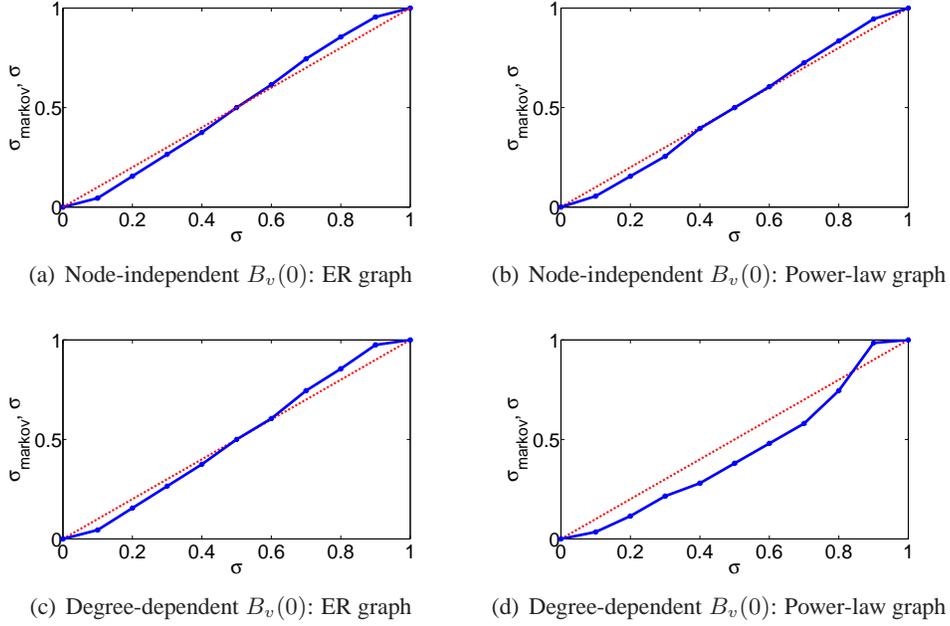

(a) Node-independent $B_v(0)$: ER graph

(b) Node-independent $B_v(0)$: Power-law graph

(c) Degree-dependent $B_v(0)$: ER graph

(d) Degree-dependent $B_v(0)$: Power-law graph

Figure 9: Threshold-drifting phenomenon: red diagonal line corresponds to $\sigma$ and blue curves correspond to $\sigma_{markov}$.

What is the cause of the threshold-drifting phenomenon? In order to answer this question, let us define $\alpha = \frac{1}{n}\sum_{v \in V} B_v(0)$, namely the average fraction of blue nodes at time $t = 0$. The probability that $k$ out of node $v$'s $\deg(v)$ neighbors are initially blue is:

$$Q(\deg(v), \alpha, k) = \binom{\deg(v)}{k}\alpha^k(1-\alpha)^{\deg(v)-k}.$$

Suppose at each time step the occupation probability approximately follows the binomial distribution. For a random node $\bar{v}$, its expected degree is $\langle \deg(v) \rangle$ and the probability that $\bar{v}$ is blue is $\nu(t) = \langle P\{\xi_v = 1\} \rangle$, with $\nu(0) = \alpha$. Now we consider the Dynamic System model. The mean of $\theta_{\bar{v},RB}(t)$ is the probability that the actual number of blue neighbors is greater than $\sigma \cdot \langle \deg(v) \rangle$. Denote this probability by $\theta_\sigma(\nu(t), \langle \deg(v) \rangle)$. Then,

$$\theta_\sigma(\nu(t), \langle \deg(v) \rangle)$$
$$= \begin{cases} \sum_{k > \nu(t) \cdot \langle \deg(v) \rangle} Q(\langle \deg(v) \rangle, \nu(t), k) & \text{if } \sigma \cdot \langle \deg(v) \rangle \text{ is no integer} \\ \sum_{k > \nu(t) \cdot \langle \deg(v) \rangle} Q(\langle \deg(v) \rangle, \nu(t), k) + \frac{1}{2} Q(\langle \deg(v) \rangle, \nu(t), \sigma \cdot \deg(v)) & \text{if } \sigma \cdot \langle \deg(v) \rangle \text{ is integer.} \end{cases}$$

Hence, we can use the following equation to approximate the Markov process model:

$$\frac{d\nu(t)}{dt} = \theta_\sigma(\nu(t), \langle \deg(v) \rangle) - \nu(t). \tag{22}$$

This one-dimension differential equation has two stable equilibria, $\nu = 0$ (i.e., all nodes are red) and $\nu = 1$ (i.e., all nodes are blue). The critical value of the initial condition between the attracting basins $\nu = 0$ and $\nu = 1$ is the non-trivial solution of $\theta_\sigma(\nu, \langle \deg(v) \rangle) - \nu = 0$, namely the solution other than 0 and 1 (which are the trivial solutions). The critical value in the Dynamic System model approximates $\sigma_{markov}$. As shown in Figure 9, $\sigma_{markov} \neq \sigma$, which explains the threshold-drifting phenomenon.

Having explained the cause of the threshold-drifting phenomenon, we suspect that the degree of threshold-drifting also depends on the average node degree (more specifically, the threshold-drifting phenomenon disappears with the average degree). To confirm/disconfirm this, we compare in Figure 10 the threshold $\sigma_{markov}$ in the Markov process model and the threshold $\sigma$ in the Dynamic System model, with respect to identical initial blue-occupation probability $B_v(0)$. In both ER and power-law graphs, we observe that $\sigma_{markov}$ asymptotically converges to $\sigma$ as the average node degree $\langle \deg(v) \rangle$ increases.

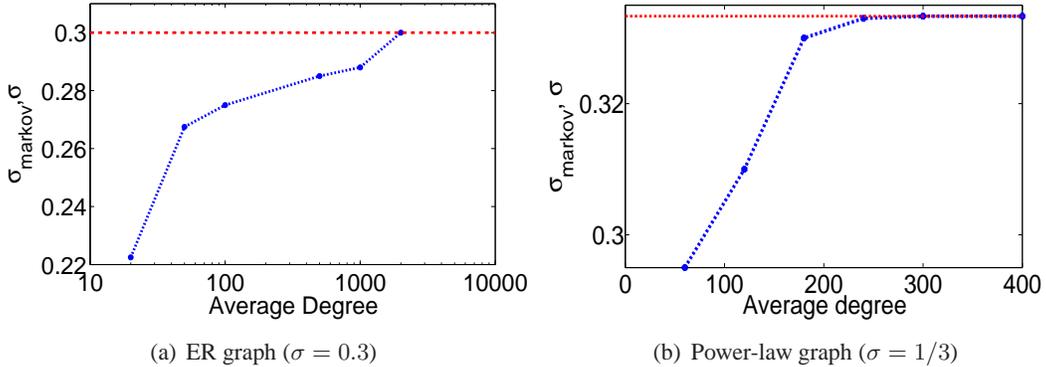

(a) ER graph ($\sigma = 0.3$)  (b) Power-law graph ($\sigma = 1/3$)

Figure 10: The greater the average node degree, the better the approximation of $\sigma_{markov}$ (blue curve) to $\sigma$ (red line).

The implication of the threshold-drifting phenomenon is that the threshold $\sigma$ may need to be adjusted in practice when $\sigma > 1/2$ (i.e., for some small $\Delta\sigma$ using $\sigma_{markov} = \sigma + \Delta\sigma$ instead). For the case $\sigma < 1/2$, adjustment is not necessary because $\sigma$ ($> \sigma_{markov}$) is sufficient for governing the dynamics toward the all-blue equilibrium (i.e., active cyber defense is effective for automatically cleaning up the network).

## 9 Conclusions

We presented the first mathematical model and characterization of active cyber defense dynamics. The analytical results give conditions under which (strategic) active cyber defense is effective, and lead to practical insights that can be adopted for decision-making and policy-making in real life.

Our study brings a range of interesting research problems, such as: How should we accommodate more sophisticated combat-power functions? How can we analyze strategic defender/attacker, including $B_v(0) \propto \deg(v)$ and possibly other scenarios, in *arbitrary* networks (rather than in the generalized random graph model)? How can we analyze the native Markov process model without using the Dynamic System approximation (while noting that the difficulty mainly lies in the nonlinearity of the combat-power functions)?

**Acknowledgement**

We thank the reviewers for their comments that helped us improve the paper. Shouhuai Xu was supported in part by ARO Grant #W911NF-12-1-0286, AFOSR MURI Grant #FA9550-08-1-0265, and NSF Grant #1111925. Wenlian Lu was jointly supported by the Marie Curie International Incoming Fellowship from the European Commission (no. FP7-PEOPLE-2011-IIF-302421), the National Natural Sciences Foundation of China (no. 61273309), the Shanghai Guidance of Science and Technology (SGST) (no. 09DZ2272900) and the Laboratory of Mathematics for Nonlinear Science, Fudan University. Any opinions, findings, and conclusions or recommendations expressed in this material are those of the author(s) and do not necessarily reflect the views of any of the funding agencies.

# A  Proof of Theorem 3

**Proof** For (i), note that equilibrium of Eq. (7) satisfies

$$f_{RB}\left(\frac{1}{\deg(v)}\sum_{u\in N_v} B_u^*\right) = B_v^*.$$

Consider a small perturbation $B(0) = B^* + \delta B$. If $B_v^* = 1$, then $\theta_{v,RB}(0) = 1$ and $B_v(t)$ increases toward 1; if $B_v^* = 0$, then $\theta_{v,RB}(0) = 0$ and $B_v(t)$ decreases toward 0. In any case, the sign of $\frac{1}{\deg(v)}\sum_{u\in N_v} B_u(t) - \sigma$ in a small time interval $[0, t_0)$ for some $t_0$ is unchanged. Let $t_1$ be the maximum time at which all the signs of $\frac{1}{\deg(v)}\sum_{u\in N_v} B_u(t) - \sigma$ are respectively the same as the signs of $\frac{1}{\deg(v)}\sum_{u\in N_v} B_u(0) - \sigma$. If $t_1$ is finite, then all signs in a small time interval starting at time $t_1$ are respectively the same as the signs at time $t_1$. This implies that $t_1 = +\infty$. So, the sign of $\frac{1}{\deg(v)}\sum_{u\in N_v} B_u(t) - \sigma$ is the same as the sign of $\frac{1}{\deg(v)}\sum_{u\in N_v} B_u(0) - \sigma$ for all $v \in V$, which implies that the system is asymptotically stable.

To see that $\bar{B}^*$ is also an equilibrium, consider the dynamic behavior of $R_v(t)$ in Eq. (7), namely $\frac{dR_v(t)}{dt} = \theta_{v,BR}(t) - R_v(t)$, where

$$\theta_{v,BR}(t) = f_{BR}\left(\frac{1}{\deg(v)}\sum_{u\in N_v} R_v(t)\right) = 1 - f_{RB}\left(\frac{1}{\deg(v)}\sum_{u\in N_v} B_v(t)\right).$$

Since $B_v(t) + R_v(t) = 1$ always holds for all $v$, $\bar{B}^*$ is an equilibrium of (11) and thus an equilibrium of (7).

To see the rates of convergence to the above equilibria, we note $\theta_{v,BR}(t) = 0$ or 1 for all $t$ and $v \in V$. Thus, (7) becomes either $\frac{dB_v(t)}{dt} = 1 - B_v(t)$, or $\frac{dB_v(t)}{dt} = -B_v(t)$. In any case, the convergence rate is $O(\exp(-t))$.

For (ii), we first note that the definition of Type I combat-power function implies $\theta_{v,RB} \in \{0, 1, \sigma\}$. Suppose at equilibrium $B^*$ that $B_{v_k}^* = \sigma$ for $v_k \in V_1 = \{v_1, \ldots, v_r\}$, where $1 \leq k \leq r$. In other words, for any $v \in V \setminus V_1$, we have $B_v^* \in \{0, 1\}$. For any $\epsilon > 0$, it is always possible to find a sufficiently small $\delta B$ from a set of positive Lebesgue measures and impose perturbation near $B^*$: $B^{**} = B^* + \delta B$ such that $\|\delta B\| < \epsilon$ and

$$\begin{cases} \frac{1}{\deg(v)}\sum_{u\in N_v} B_u^{**} > \sigma & \text{if } v \in V_1 \text{ or } B_v^* = 1 \\ \frac{1}{\deg(v)}\sum_{u\in N_v} B_u^{**} < \sigma & \text{if } B_v^* = 0. \end{cases}$$

By treating $B^{**}$ as the initial security state at time $t = 0$, there exists a time interval $[0, t_0)$ such that

- for any node $v$ with $B_v^* = \sigma$, $B_v(t)$ monotonically strictly increases toward 1 for $t \in [0, t_0)$;
- for any node $v$ with $B_v^* = 1$, we have $B_v(t) = 1$ for $t \in [0, t_0)$;
- for any node $v$ with $B_v^* = 0$, $B_v(t)$ does not decrease for $t \in [0, t_0)$.

Since for any $v$ with $B_v^{**} = \sigma$, we have $B_v(t) \to 1$ as $t \to \infty$, $B^*$ with $B_v^* = \sigma$ for some $v$ is unstable. ∎

# B  Proof of Theorem 4

**Proof** From condition $\underline{\lim}_{n\to\infty}\phi(n) > \sigma$, for almost every sequence of $\phi(n)$ we can pick some $\mu > \mu' > 0$ such that $\phi(n) > \sigma + \mu > \sigma + \mu'$ for sufficiently large $n$. Recall random variable $\chi_v(S)$:

$$\chi_v(S) = \begin{cases} 1 & v \in S \\ 0 & v \notin S. \end{cases}$$

Note that $\mathsf{P}(\chi_v(S) = 1) = B_v(0)$. Let $\zeta_{vu}$ be a random variable indicating the link between (from) node $u$ and (to) node $v$, namely

$$\zeta_{vu} = \begin{cases} 1 & (u,v) \in E(n) \\ 0 & (u,v) \notin E(n). \end{cases}$$

According to Eq. (10), we have

$$\mathsf{P}(\zeta_{vu} = 1) = p_{vu}(n) = \frac{d_v(n)d_u(n)}{\sum_{k \in V(n)} d_k(n)}.$$

Since we assumed that the $B_v(0)$'s are independent of each other and also independent of the linking of edges in $G(n)$, $\zeta_{vu}$ and $\chi_u(S)$ are independent with respect to $u$. Our goal is to estimate the probability of event $A_v$ as defined by

$$A_v = \left\{ \frac{1}{\deg(v)} \sum_{u \in N_v} B_u(0) < \sigma \right\} = \left\{ \frac{1}{\deg(v)} \sum_{u \in V(n)} B_u(0) \cdot \zeta_{vu} < \sigma \right\},$$

namely $\mathsf{P}(A_v) = \mathsf{P}(\sum_{u \in V(n)} \zeta_{vu} \cdot B_u(0) < \sigma \cdot \deg(v))$.

Note that random variables $\zeta_{vu}$ for all $u \in V(n)$ are independent of each other. Its expectation is $\mathsf{E}(\zeta_{vu}) = p_{vu}(n)$ and variance is $\mathsf{Var}(\zeta_{vu}) = p_{vu}(n)(1 - p_{vu}(n))$. Because $\mathsf{E}(\chi_u(S) \cdot \zeta_{vu}) = B_u(0)p_{vu}(n)$ and the random variable has only two states, we have

$$\mathsf{P}(A_v)$$
$$= \mathsf{P}\left( \frac{1}{\sqrt{\sum_{u \in V(n)} \mathsf{Var}(\zeta_{vu})B_u^2(0)}} \sum_{u \in V(n)} [\zeta_{vu}B_u(0) - \mathsf{E}(\zeta_{vu})B_u(0)] < \frac{\sigma \cdot \deg(v) - \sum_{u \in V(n)} \mathsf{E}(\zeta_{vu} \cdot \chi_u(S))}{\sqrt{\sum_{u \in V(n)} \mathsf{Var}(\zeta_{vu} \cdot \chi_u(S))}} \right)$$
$$= \mathsf{P}\left( \frac{1}{\sqrt{\sum_{u \in V(n)} \mathsf{Var}(\zeta_{vu})B_u^2(0)}} \sum_{u \in V(n)} [\zeta_{vu} \cdot B_u(0) - B_u(0) \cdot p_{vu}(n)] < \frac{d_v}{s_{n,v}} \left[ \sigma - \frac{\sum_{u \in V(n)} \chi_u(S) \cdot \deg(u)}{\sum_{p \in V(n)} \deg(p)} \right] \right.$$
$$+ \sigma \frac{\deg(v) - d_v}{s_{n,v}} + \frac{d_v}{s_{n,v}} \left[ \frac{\sum_{u \in V(n)} \chi_u(S) \deg(u)}{\sum_{p \in V(n)} \deg(p)} \frac{\sum_{v \in V(n)} d_v - \sum_{v \in V(n)} \deg(v)}{\sum_{v \in V(n)} d(v)} \right.$$
$$\left. \left. + \frac{\sum_{u \in V(n)} \deg(u)\chi_u(S) - \sum_{u \in V(n)} d_u B_u(0)}{\sum_{p \in V(n)} d_p} \right] \right)$$

where $s_{n,v}^2$ is defined in Eq. (12). Since $\mathsf{Var}[\zeta_{v,u}B_u(0)] = B_u(0)^2 p_{vu}(1 - p_{vu})$, we have $\sum_{u \in V(n)} \mathsf{Var}[\zeta_{v,u}B_u(0)] = s_{n,v}^2$.

Note that assumption (i) implies

$$\lim_{n \to \infty} \frac{1}{s_{n,v}^3} \sum_{u \in V(n)} \left\{ \mathsf{E} |\zeta_{v,u}B_u(0) - \mathsf{E}(\zeta_{v,u}B_u(0))|^3 \right\} = \lim_{n \to \infty} \frac{q_{n,v}}{s_{n,v}^3} = 0.$$

with $q_{n,v}$ defined in Eq. (13). This guarantees the Lyapunov condition in the Central Limit Theorem (with $\delta = 1$) [41]. So, as $n \to \infty$,

$$\frac{1}{\sqrt{s_{n,v}^2}} \sum_{u \in V(n)} [\zeta_{v,u}B_u(0) - B_u(0)p_{vu}] \to N(0,1) \tag{23}$$

in distribution uniformly, as $n \to \infty$. We call this asymptotic normal random variable $\phi_v$.

In addition, we observe that

$$\sigma \frac{\deg(v) - d_v}{s_{n,v}} = \sigma \frac{d_v}{s_{n,v}} \frac{\deg(v) - d_v}{d_v} = \sigma \frac{d_v}{s_{n,v}} \frac{\deg(v) - d_v}{w_{n,v}} \frac{w_{n,v}}{d_v} \to o(1) \frac{d_v}{s_{n,v}} \quad (24)$$

with $w_{n,v}$ defined in Eq. (14), with probability 1, because the term $\frac{\deg(v)-d_v}{w_{n,v}}$ converges to the standard Gaussian random variable owing to the Lyapunov central limit theorem where the Lyapunov condition is guaranteed by assumption (ii), noting $g_{n,v}$, defined in Eq. (15), denoting the third order moment of $\zeta_{v,u}$ for all $u \in V(n)$, and because

$$\frac{w_{n,v}}{d_v} \le \frac{1}{\sqrt{d_v}} \to 0$$

as $n \to \infty$, owing to assumption (iii). Furthermore, we observe that

$$\frac{\sum_{v \in V(n)} d_v - \sum_{v \in V(n)} \deg(v)}{\sum_{v \in V(n)} d(v)} = \frac{\sum_{v \in V(n)} d_v - \sum_{v \in V(n)} \deg(v)}{\sqrt{\sum_{v \in V(n)} w_{n,v}^2}} \frac{\sqrt{\sum_{v \in V(n)} w_{n,v}^2}}{\sum_{v \in V(n)} d(v)} \to 0 \quad (25)$$

almost surely, because $\frac{\sum_{v \in V(n)} d_v - \sum_{v \in V(n)} \deg(v)}{\sqrt{\sum_{v \in V(n)} w_{n,v}^2}}$ converges to the standard Gaussian random variable, owing to the Lyapunov central limit theorem where the Lyapunov condition is guaranteed by assumption (iv), and

$$\frac{\sqrt{\sum_{v \in V(n)} w_{n,v}^2}}{\sum_{v \in V(n)} d(v)} \le \frac{1}{\sqrt{\sum_{v \in V(n)} d_v}} \to 0,$$

owing to assumption (iii). We further observe that

$$\frac{\sum_{u \in V(n)} \deg(u) \chi_u(S) - \sum_{u \in V(n)} d_u B_u(0)}{\sum_{p \in V(n)} d_p}$$

$$= \frac{\sum_{u \in V(n)} \deg(u) \chi_u(S) - \sum_{u \in V(n)} d_u B_u(0)}{\sqrt{\sum_{v \in v(n)} s_{n,v}^2}} \frac{\sqrt{\sum_{v \in v(n)} s_{n,v}^2}}{\sum_{p \in V(n)} d_p} \to 0 \quad (26)$$

almost surely, because

$$\frac{\sum_{u \in V(n)} \deg(u) \chi_u(S) - \sum_{u \in V(n)} d_u B_u(0)}{\sqrt{\sum_{v \in v(n)} s_{n,v}^2}}$$

converges to the standard Gaussian random variable, owing to the Lyapunov central limit theorem where the Lyapunov condition is guaranteed by assumption (v), and

$$\frac{\sqrt{\sum_{v \in v(n)} s_{n,v}^2}}{\sum_{p \in V(n)} d_p} \le \frac{1}{\sqrt{\sum_{p \in V(n)} d_p}}$$

owing to assumption (iii). Combining (23), (24), (25) and (26) with the fact

$$\frac{\sum_{u \in V(n)} \chi_u(S) \deg(u)}{\sum_{p \in V(n)} \deg(p)} \le 1,$$

we conclude that there exists a random variable $\epsilon_{n,v}$ that converges to zero uniformly with probability 1 such that

$$P(A_v) = P\left(\frac{1}{\sqrt{\sum_{u \in V(n)} \text{Var}(\zeta_{vu}) B_u^2(0)}} \sum_{u \in V(n)} [\zeta_{vu} B_u(0) - E(\zeta_{vu}) B_u(0))] < \frac{d_v}{s_{n,v}}\left(\sigma - \phi(n) + \epsilon_{n,v}\right)\right).$$

Finally, we observe that $\sigma - \phi(n) \leq -\mu$ holds with probability 1. This inequality, together with the convergence rate in the Central Limit Theorem [46], implies

$$\left|P\left(A_v \middle| \eta \geq \sigma + \mu\right) - \Phi(t_n(v))\right| \leq C \frac{q_{n,v}/s_{n,v}^3}{(1 + |t_n(v)|^3)},$$

where $t_n(v) = -\mu' \frac{\deg(v)}{s_{n,v}}$, noting $\mu' < \mu$, for sufficiently large $n$, $\Phi(\cdot)$ is the probability function of the standard normal distribution, and $C$ is a universal constant.

Since

$$\Phi(t_n(v)) = \frac{2}{\sqrt{\pi}} \int_{-\infty}^{t_n(v)} \exp\left(-\frac{y^2}{2}\right) dy,$$

and

$$\int_{-\infty}^{x} \exp(-y^2/2) dy \leq \exp(-x^2/2)/(-x) \text{ for all } x \leq 0,$$

we have

$$\sum_{v \in V(n)} \Phi(t_n(v)) < \frac{2n}{\sqrt{\pi}} \frac{\exp\left[-(\min_v t_n(v))^2/2\right]}{\min_v t_n(v)}. \tag{27}$$

Under assumption (iii), the supra-limit of the logarithm of the right-hand side of Eq. (27) becomes:

$$\overline{\lim}_{n \to \infty} \left\{\ln\left(\frac{2}{\sqrt{\pi}}\right) + \ln(n) - [\min_v t_n(v)]^2/2 - \ln[\min_v t_n(v)]\right\} = -\infty.$$

This implies $\sum_{v \in V(n)} \Phi(t_n(v)) \to 0$ as $n \to \infty$. In addition, we observe that

$$\sum_{v \in V(n)} \frac{C q_{n,v}}{1 + |t_{n,v}|^3} \leq C \sum_{v \in V(n)} \frac{q_{n,v}}{d_v^3} \leq C \sum_{v \in V(n)} \frac{1}{d_v^2}$$

converges to zero owing to assumption (vi).

Putting the above together, we have

$$\lim_{n \to \infty} P\left(\bigcup_{v \in V(n)} A_v\right) \leq \overline{\lim}_{n \to \infty} \sum_{v \in V(n)} P(A_v)$$

$$\leq \overline{\lim}_{n \to \infty} \sum_{v \in V(n)} \Phi(t_n(v)) + C \overline{\lim}_{n \to \infty} \sum_{v \in V(n)} \frac{C q_{n,v}}{1 + |t_n(v)|^3}$$

$$= 0$$

By applying Theorem 1, for each event not belonging to $\bigcup_v A_v$, we have $\lim_{t \to \infty} B_v(t) = 1$ for all $v \in V(n)$. This proves the first part of the theorem.

Analogously, we can prove the second part. This completes the proof. ∎

# C Proof of Lemma 2

**Proof** We only prove part (i) because part (ii) can be proved analogously. In order to simplify the presentation, let $Y_v(t) = \frac{1}{\deg(v)} \sum_{u \in N_v} B_u(t)$ be the average portion of $v$'s blue neighbors at time $t$.

First, we need to show that $Y_v(t) > \tau$ holds for all $v \in V$ and $t \geq 0$. For this purpose, we let $\tau^* > \tau$ such that $Y_v(0) > \tau^*$ holds for all $v \in V$, and show that $Y_v(t) > \tau^*$ holds for all $t \geq 0$. We observe that $Y_v(t) > \tau^*$ holds in a small time interval starting at time $t = 0$ because of the continuity of the $B_v(t)$'s with respect to $t$. Let $t_1$ be the first time at which $\min_{v \in V} Y_v(t) = \tau^*$, namely

$$t_1 = \inf\left\{ \mathsf{t} : \min_{v \in V} Y_v(t) > \tau^* \text{ for all } t \in [0, \mathsf{t}) \in V \right\}.$$

We show $t_1 = +\infty$ as follows.

Suppose $t_1 < +\infty$. We claim that $\min_{v \in V} Y_v(t)$ is non-increasing in an interval starting at time $t_1$; otherwise, $\frac{d(\min_{v \in V} Y_v(t))}{dt} > 0$ in a small interval starting at time $t_1$ and $\min_v Y_v(t) > \tau^*$ in the small interval, which contradicts the definition of $t_1$.

Let $V^* = \arg\min_{v \in V} Y_v(t_1)$. For each $v' \in V^*$, we have

$$\frac{d}{dt}\left[ \frac{1}{\deg(v')} \sum_{u \in N_{v'}} B_u(t) \right]\bigg|_{t=t_1}$$

$$= \frac{1}{\deg(v')} \sum_{u \in N_{v'}} \frac{dB_u(t)}{dt}\bigg|_{t=t_1}$$

$$= \frac{1}{\deg(v')} \sum_{u \in N_{v'}} \left[ f_{RB}\left( \frac{1}{\deg(u)} \sum_{w \in N_u} B_w(t_1) \right) - B_u(t_1) \right]$$

$$\geq \frac{1}{\deg(v')} \sum_{u \in N_{v'}} f_{RB}(\tau^*) - \tau^*$$

$$= f_{RB}(\tau^*) - \tau^* > 0$$

owing to $\tau^* > \tau$. Hence, $\min_v Y_v(t)$ is strictly increasing in an interval starting at time $t_1$. This contradicts that $\min_{v \in V} Y_v(t)$ is non-increasing in an interval starting at time $t_1$. The contradiction was caused by the assumption $t_1 < +\infty$. Therefore, we have $t_1 = +\infty$.

Second, we need to show that $\min_{v \in V} B_v(t)$ increases monotonically. Let $V_t = \{u : B_u(t) = \arg\min_v B_v(t)\}$, which may not be a singlet. For $t = 0$, the given initial condition $\frac{1}{\deg(v)} \sum_{u \in N_v} B_u(0) > \tau$ for all $v \in V$ implies that for each $v_* \in V_0$, we have

$$\frac{dB_{v_*}(t)}{dt}\bigg|_{t=0} = f_{RB}\left( \frac{1}{\deg(v_*)} \sum_{u \in N_{v_*}} B_u(0) \right) - B_{v_*}(0)$$

$$> \frac{1}{\deg(v_*)} \sum_{u \in N_{v_*}} B_u(0) - B_{v_*}(0) \geq 0$$

because $f_{RB}(s) > s$ for $s > \tau$. This means that $\min_v B_v(t)$ strictly increases in a small time interval starting at $t = 0$.

Let $t_2$ be the maximum time that $\min_v B_v(t)$ keeps strictly increasing, namely

$$t_2 = \sup\{\mathsf{t} : \min_v B_v(t) \text{ strictly increases in } [0, \mathsf{t})\}.$$

We now show that $t_2 = +\infty$. Suppose $t_2 < +\infty$, meaning that $\min_v B_v(t)$ is not strictly increasing at $t = t_2$. However, for each $v_2 \in V_{t_2}$, we have

$$\begin{aligned}\left.\frac{dB_{v_2}(t)}{dt}\right|_{t=t_2} &= f_{RB}\left(\frac{1}{\deg(v_2)}\sum_{u \in N_{v_2}} B_u(t_2)\right) - B_{v_2}(t_2) \\ &> \frac{1}{\deg(v_2)}\sum_{u \in N_{v_2}} B_u(t_2) - B_{v_2}(t_2) \geq 0\end{aligned}$$

because $\frac{1}{\deg(v_2)}\sum_{u \in N_{v_2}} B_u(t_2) > \tau$ and $f_{RB}(s) > s$ for all $s > \tau$. This implies that $\min_v B_v(t)$ strictly increases at $t = t_2$, which contradicts the definition of $t_2$. Therefore, $t_2 = +\infty$, namely $\min_v B_v(t)$ strictly increases for all $t \geq 0$. ∎

## D  Proof of Theorem 5

**Proof**  We prove the first part as the second part can be proved analogously. Lemma 2 shows that $\min_v B_v(t)$ monotonically increases, meaning that $\lim_{t \to \infty} \min_v B_v(t)$ exists. In order to show $\lim_{t \to \infty} B_v(t) = 1$ for all $v \in V$, it suffices to show $\lim_{t \to \infty} \min_v B_v(t) = 1$. Suppose $\lim_{t \to \infty} \min_v B_v(t) < 1$. There are two cases, but both cause contradictions as we elaborate below. Therefore, we have $\lim_{t \to \infty} \min_v B_v(t) = 1$.

**Case 1:** $\tau < \lim_{t \to \infty} \min_v B_v(t) < 1$.

There exist $\tau < \tau_1 < \tau_2 < 1$ and $T > 0$ such that $\tau_1 \leq \min_v B_v(t) \leq \tau_2$ for all $t \geq T$. Since $f_{RB}(x) - x > 0$ for all $x \in [\tau_1, \tau_2]$ and is continuous, we can find some $\delta > 0$ such that $f_{RB}(x) - x > \delta$ for all $x \in [\tau_1, \tau_2]$. Let $V_t$ be the index set of $\arg\min_v B_v(t)$. For each $v_* \in V_t$, we have

$$\begin{aligned}\frac{dB_{v_*}(t)}{dt} &= f_{RB}\left(\frac{1}{\deg(v_*)}\sum_{u \in N_{v_*}} B_u(t)\right) - B_{v_*}(t) \\ &\geq f_{RB}(B_{v_*}(t)) - B_{v_*}(t) > \delta\end{aligned} \qquad (28)$$

for all $t > T$. This leads to

$$\min_v B_v(t) > \min_v B_v(T) + \delta(t - T).$$

Since $\min_v B_v(T) + \delta(t - T) \to +\infty$ as $t \to \infty$, this contradicts $B_v(t) \leq 1$.

**Case 2:** $\lim_{t \to \infty} \min_v B_v(t) \leq \tau$.

Let $V_t$ be the index set of $\arg\min_v B_v(t)$. Since $\frac{1}{\deg(v)}\sum_{u \in N_v} B_u(t) > \tau$ for all $v$ and $t$, there exist $T' > 0$ and $\delta' > 0$ such that for each $v_* \in V_t$,

$$f_{RB}\left(\frac{1}{\deg(v_*)}\sum_{u \in N_{v_*}} B_u(t)\right) - B_{v_*}(t) > \delta'$$

holds for all $t > T'$. By the same argument as in **Case 1**, we can show $\lim_{t \to \infty} \min_v B_v(t) = +\infty$, which contradicts $B_v(t) \leq 1$. ∎

# E  Proof of Theorem 7

**Proof**  Part (i) can be seen by considering any perturbation near each equilibrium $B^*$. For these equilibria, we can use linearization to analyze the convergence rates. Let $B(t) = ([B_v(t)]_{v \in V})^\top$, $A$ be the adjacency matrix of $G$, $D = \text{diag}\,([\deg(v)]_{v=1}^n)$, $\mathbf{1} = [1, \ldots, 1]^\top$, $\mathbf{0} = [0, \ldots, 0]^\top$, $\delta B$ be the variation of $B(t)$ near $\mathbf{1}$ or $\mathbf{0}$, $I_n$ denote the $n$-dimension identity matrix, $z = 1$ indicate that we are considering the convergence rate of stable equilibrium $\mathbf{1}$, and $z = 0$ indicate that we are considering the convergence rate of stable equilibrium $\mathbf{0}$. Then, linearization leads to

$$\frac{d\delta B(t)}{dt} = \left[ f'_{RB}(z) D^{-1} A - I_n \right] \delta B.$$

The convergence rate is estimated by the largest real part of all eigenvalues of matrix $f'_{RB}(z) D^{-1} A - I_n$. Since the largest eigenvalue of $D^{-1} A$ equals 1, the convergence rate is estimated as $O(\exp[(f'_{RB}(z) - 1)t])$ for both $z = 0$ and $z = 1$.

For proving part (ii), suppose at equilibrium $B^*$ that $B^*_{v_k} = \tau$ for $v_k \in V_1 = \{v_1, \ldots, v_r\}$, where $1 \leq k \leq r \leq n$. In other words, for any $v \in V \setminus V_1$, $B^*_v \in \{0, 1\}$. For any $\epsilon > 0$, it is always possible to find a sufficiently small $\delta B$ from a set of positive Lebesgue measures and impose a perturbation near $B^*$ while satisfying the following: $B^{**} = B^* + \delta B$ such that $\|\delta B\| < \epsilon$ and

$$\begin{cases} \frac{1}{\deg(v)} \sum_{u \in N_v} B^{**}_u > \tau & \text{if } B^*_v \geq \tau \\ \frac{1}{\deg(v)} \sum_{u \in N_v} B^{**}_u < \tau & \text{if } B^*_v < \tau. \end{cases} \quad (29)$$

Let us treat $B^{**}$ as the initial security state at time $t = 0$. For any node $v$ with $B^*_v \geq \tau$, we have

$$\left. \frac{dB_v(t)}{dt} \right|_{t=0} = f_{RB}\left( \frac{1}{\deg(v)} \sum_{u \in N_v} B^{**}_u \right) - B^*_v$$

$$> \frac{1}{\deg(v)} \sum_{u \in N_v} B^{**}_u - B^{**}_v \geq 0.$$

This means that there is a time interval $[0, t_0)$ in which for any node $v$ with $B^*_v \geq \tau$, $B_v(t)$ monotonically strictly increases. For any node $v$ with $B^*_v < \tau$, we have

$$\left. \frac{dB_v(t)}{dt} \right|_{t=0} = f_{RB}\left( \frac{1}{\deg(v)} \sum_{u \in N_v} B^{**}_u \right) - B^*_v$$

$$< \frac{1}{\deg(v)} \sum_{u \in N_v} B^{**}_u - B^{**}_v \leq 0,$$

which means that the corresponding $B_v(t)$'s strictly decrease for a small time interval $t \in [0, t_0)$. In summary, for any small perturbation with (29) as the initial security state, $B_v(t)$ leaves the equilibrium. Therefore, $B^*$ with $B^*_v = \tau$ for some $v$ is unstable. ∎